\documentclass[12pt,preprint]{aastex}
\usepackage{graphicx}
\begin{document}

\title{The Relativistic Jet-Accretion Flow-Wind Connection in Mrk\,231}
\author{Cormac Reynolds\altaffilmark{1}, Brian Punsly\altaffilmark{2}, Giovanni Miniutti\altaffilmark{3}, Christopher P. O'Dea\altaffilmark{4,5}, Natasha Hurley-Walker\altaffilmark{6}}
\altaffiltext{1}{CSIRO Astronomy and Space Science, Kensington, WA 6151, Australia}
 \altaffiltext{2}{1415 Granvia Altamira, Palos
Verdes Estates CA, USA 90274 and ICRANet, Piazza della Repubblica 10
Pescara 65100, Italy, brian.punsly1@verizon.net}
\altaffiltext{3}{Centro de Astrobiologia (CAB) ESA - European Space
Astronomy Center (ESAC)} \altaffiltext{4}{Department of Physics and
Astronomy, University of Manitoba, Winnipeg, MB R3T 2N2
Canada}\altaffiltext{5}{School of Physics \& Astronomy, Rochester
Institute of Technology, Rochester, NY 14623, USA} \altaffiltext{6}
{ICRAR-Curtin University, GPO Box U1987, Perth, Western Australia,
6102, Australia}

\begin{abstract}
Long term radio monitoring of the broad absorption line quasar,
Mrk\,231, at 17.6~GHz detected a strong flare in 2015. This
triggered four epochs of Very Long Baseline Array (VLBA)
observations from 8.4 GHz to 43 GHz as well as three epochs of X-ray
observations with NuSTAR and two with XMM over a 15 week period. Two
ejected components were detected by the VLBA observations. A
conservative lower bound on the apparent speed of the first ejection
is attained by assuming that it was ejected when the flare began,
$v_{\rm{app}}>3.15$c. Serendipitous far UV Hubble Space Telescope
observations combined with our long term radio monitoring seem to
indicate that episodes of relativistic ejections suppress flux that
is emitted at wavelengths shortward of the peak of the far UV
spectral energy distribution, similar to what has been observed in
radio loud quasars. Episodes of strong jet production also seem to
suppress the high ionization broad absorption line wind seen in weak
jet states. We found a statistically significant increase ($\sim
25\%$) of the 3-12 keV flux during the radio flare relative to a
quiescent radio state. This is explained by an ultra-fast ($\sim
0.06$c) X-ray absorbing photo-ionized wind that is significantly
detected only in the low radio state (similar to Galactic black
holes). Mrk~231 is becoming more radio loud. We found that the
putative parsec scale radio lobe doubled in brightness in 9 years.
Furthermore, large flares are more frequent with 3 major flares
occurring at $\sim 2$ year intervals.
\end{abstract}
\keywords{quasars: absorption lines --- galaxies: jets
--- quasars: general --- accretion, accretion disks --- black hole physics}

\section{Introduction} Mrk\,231, at a redshift of $z = 0.042$, is
a nearby radio quiet quasar (RQQ) with a relativistic jet
\citep{rey09}. The jet is extremely powerful for a RQQ during flare
states with a kinetic luminosity estimated to be $\sim 3\times
10^{43}$ ergs~$\rm{s}^{-1}$ for previous flares. Consequently, we
have been monitoring the radio behavior at $\sim 20$~GHz for the
last 7 years. Since 2011, we have detected 3 large blazar-like
flares (see \citet{rey13}) with flux densities $\geq 200$~mJy at
$\sim 20$~GHz. The last of these major flares initiated a target of
opportunity four-epoch VLBA (Very Long Baseline Array) observation.
We were also afforded director's discretionary time for three
observations with the NuSTAR X-ray telescope and two observations
with the XMM X-ray telescope.
\par Mrk\,231 is a unique laboratory for studying the multiple
aspects that are associated with the quasar phenomenon. It is one of
a small number of RQQs that have clearly exhibited episodes of
relativistic jet formation \citep{bru00,blu03}. It also supports
broad absorption line (BAL) winds. We use the original definition of
BAL quasars (BALQSOs) as quasars with UV absorbing gas that is blue
shifted at least 5,000 km/s relative to the QSO rest frame and
displaying a spread in velocity of at least 2,000~km~s$^{-1}$,
\citep{wey91}. This definition excludes the so-called
``mini-BALQSOs," with the BALNicity index = 0 \citep{wey97}. This
definition is preferred here since mini-BALQSOs tend to resemble
non-BALQSOs more than BALQSOs in many spectral and broadband
properties \citep{pun06,zha10, bru13, hay13}. A nonzero BALNicity
index is extraordinary considering that the existence of large scale
jets and BAL winds are almost mutually exclusive. The propensity for
suppressed large scale emission increases strongly with BALnicity
index \citep{bec00,bec01}. Thus, episodes of relativistic jet
production in Mrk\,231 allow us to explore the interplay between BAL
winds and jets. Mrk\,231 is exceptional in that it tends to
vacillate between radio quiet and radio loud behaviors. In
particular, in this study we find states of weak jet activity have
UV characteristics that differ from those of the states of high jet
activity. Based on limited data, the weak jet state shows evidence
of high ionization BAL winds, as noted above, a property associated
with radio quiet quasars. In the states of high jet activity, the
evidence of a high ionization BAL wind disappears and the spectral
energy distribution steepens beyond its peak in the far UV, a
property common to radio loud quasars \citep{tel02,pun15}. This
transition between high jet activity and low jet activity occurs on
very short time scales in some Galactic black holes
\citep{kle02,pra10}. Consequently, these stellar mass objects form
the basis of our physical understanding of the state change (both
back and forth) between a jetted black hole accretion system to a
non-jetted black hole accretion system. It is not clear that the
trends observed in Galactic black holes extrapolate many orders of
magnitude in both time and space to quasars. This was the motivation
for our previous high frequency (15~GHz, 22~GHz and 43~GHz) VLBA
observations \citep{rey09}. Our previous VLBA observation detected a
strong 22~GHz flare that emerged from the core between epochs
2006.07 and 2006.32 ($>150\%$ increase in less than 3 months).
However, we learned that the putative ejecta faded very rapidly and
required both high resolution (43~GHz) and high sensitivity. Thus,
we recognised the need to observe a strong flare that was still
growing and to improve the sensitivity by choosing a higher data
sampling rate, 2 Gbps (as opposed to the 256 Mbps that was
previously available). This observation challenged the limits of the
VLBA and our efforts were designed to maximize the likelihood of
successful phase-referencing and the subsequent self-calibration of
the weak target source. In order to determine the initiation of a
Target of Opportunity (ToO) for our VLBA observations, we began
monitoring Mrk\,231 at 22~GHz in 2009, switching to 17.6~GHz in
2012. The details of our monitoring program and the discovery of a
large flare were previously reported \citep{rey13}.

\par The paper is organized as follows. Section~2 describes the observational details of
our monitoring. In Section 3, we describe the VLBA observations that
were triggered on 2015 March 31. There were four observations at 8.4, 15.2, 22.2
and 43~GHz that were spread out over 15 weeks. In Section~4, we describe Hubble
Space Telescope (HST) observations of the far UV spectrum of Mrk\,231 that
serendipitously align with three epochs of our radio monitoring. In Section 5,
we discuss the X-ray telescope observations that were granted after our VLBA
ToO, three NuSTAR observations and two XMM observations.
Throughout this paper, we adopt the following cosmological
parameters: $H_{0}$=71 km s$^{-1}$ Mpc$^{-1}$,
$\Omega_{\Lambda}=0.73$ and $\Omega_{m}=0.27$.

\begin{figure}
\begin{center}
\includegraphics[width= 0.8\textwidth]{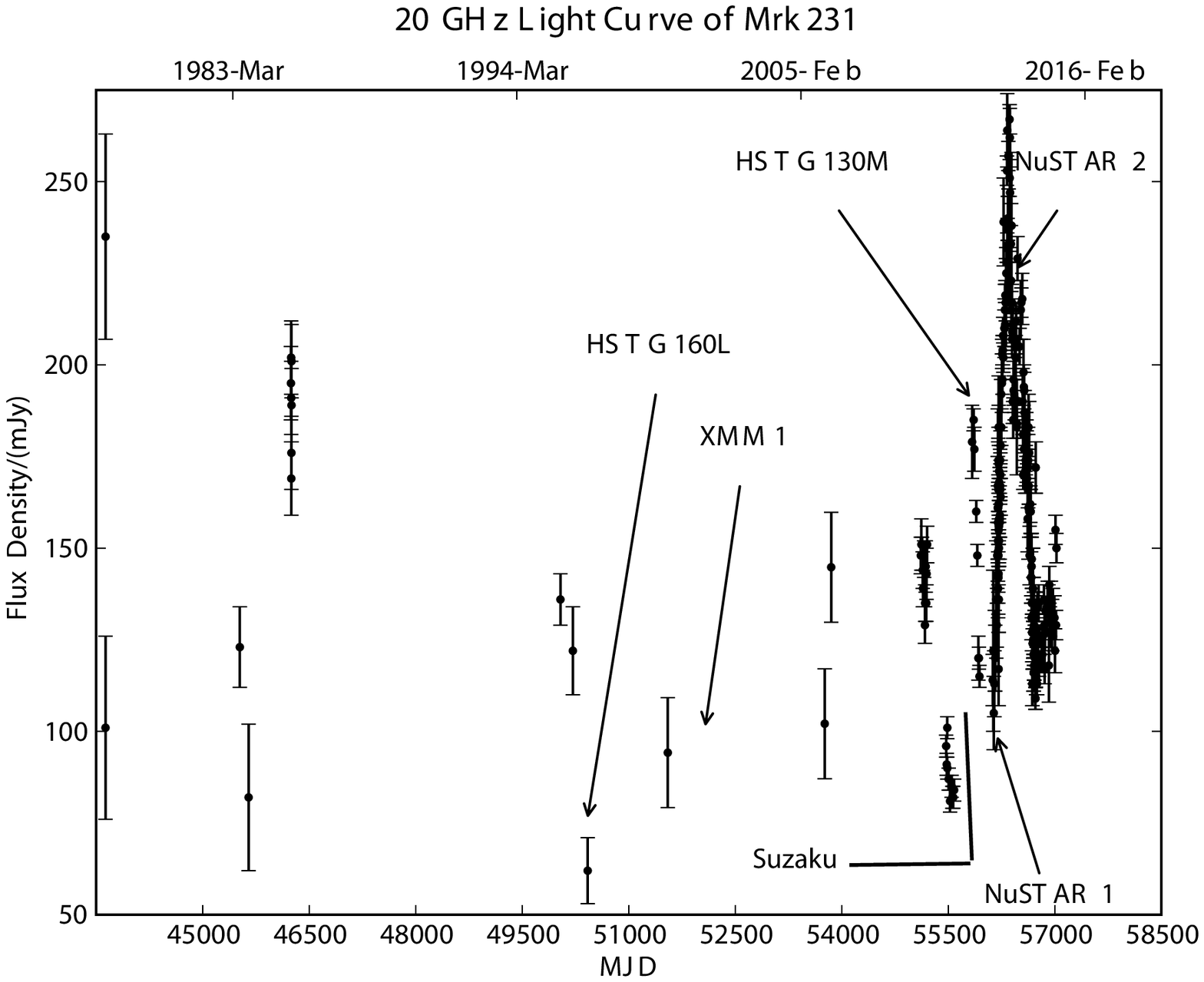}
\includegraphics[width= 0.8\textwidth]{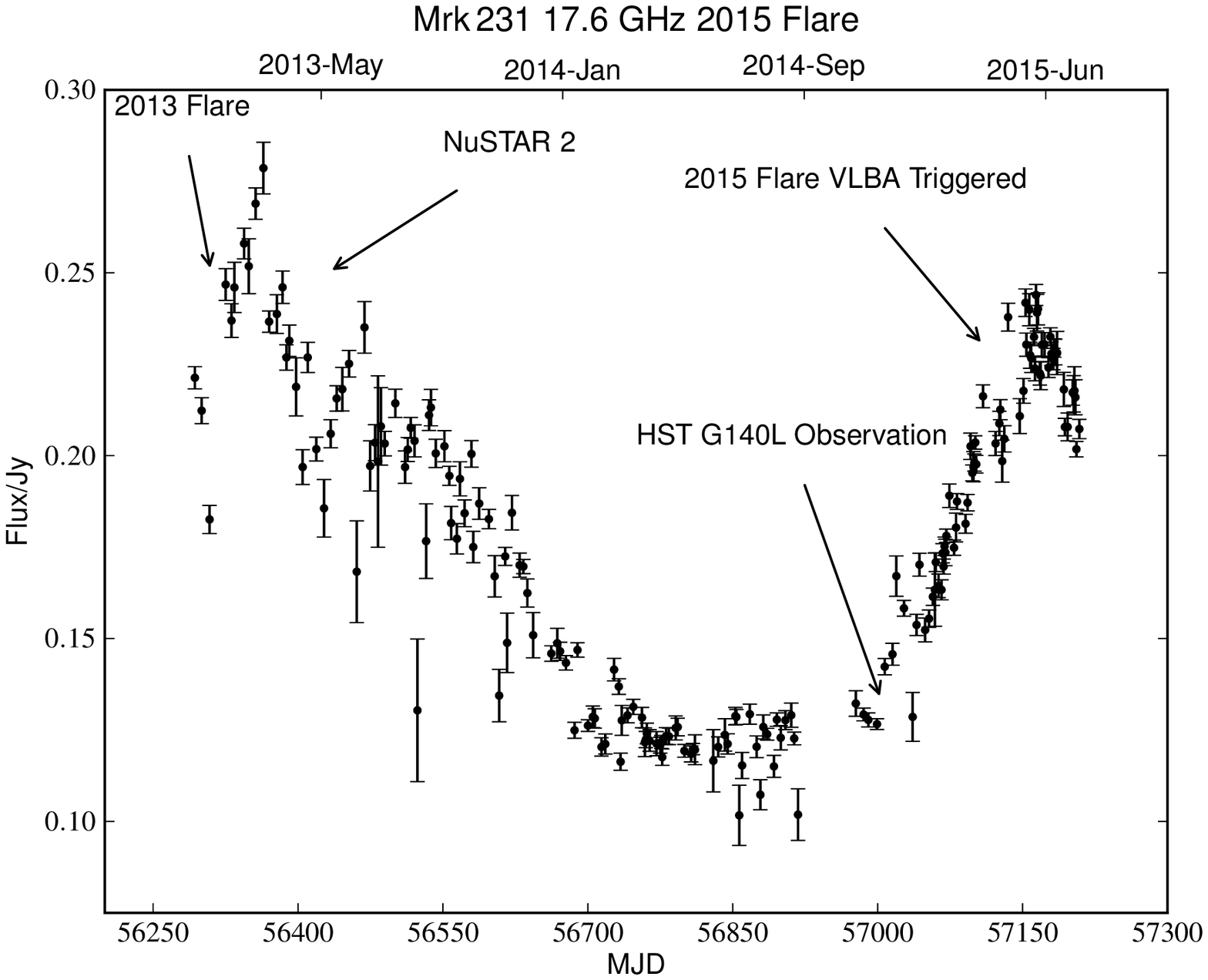}

\caption{The top frame is the historic $\sim20$~GHz light curve from
\citet{rey13} with the timing of observations in other energy bands
that are discussed in this paper superimposed. The bottom frame is
an AMI 17.6~GHz light curve showing the initiation of the radio
flare discussed in this article. }
\end{center}
\end{figure}

\section{The Radio Monitoring}
\citet{rey13} reported the methods of our radio monitoring program.
In summary, the program utilized long term monitoring with the VLA
(Very Large Array) and EVLA (Expanded Very Large Array) at 22~GHz
from 2009 to 2012 and from 2013 to 2015 AMI \citep[Arcminute
Microkelvin Imager;][]{zwa08} at 13.5 -- 18~GHz \footnote{The
Arcminute Microkelvin Imager consists of two radio interferometric
arrays located in the Mullard Radio Astronomical Observatory,
Cambridge, UK \citep{zwa08}. Observations occur between 13.9 and
18.2~GHz in six frequency channels. The Small Array (AMI-SA)
consists of ten 3.6~m diameter dishes with a maximum baseline of
20~m, with an angular resolution of 3\arcmin, while the Large Array
(AMI-LA) comprises eight 12.6~m diameter dishes with a maximum
baseline of 110~m, giving an angular resolution of 0\farcm5.}. The
long term monitoring data from \citet{rey13} are plotted in the top
frame of Figure 1. This light curve incorporates historical data
both from the literature \citep{mcc78,ede87,ulv99,ulv00,rey09} and
from the VLA public archive (project codes AB783, AN030, AU015). We
have superimposed the times of relevant observations that we will
reference in the following. The notable features of the historical
light curve are the major flare of 2013 that was monitored with AMI
and analyzed previously \citep{rey13}. It was shown that this is
similar to blazar flares (rise time, decay and relative magnitude),
but on a smaller scale. The unexpected steady rise of flux failed to
meet our trigger criteria (which were then revised) and we missed
this opportunity to study a flare in progress. The other important
feature is the HST Faint Object Spectrograph (FOS) observation with
the G160L grating on 1996 November 21, just 14 days before the
historically low VLA flux density measurement \citep{ulv99}. This
fortuitous circumstance will form an important part of our analysis
in Section~4. Another fortuitous circumstance is the HST Cosmic
Origins Spectrograph (COS) observation with the G130M grating on
2011 October 15 reported in \citet{vei13} that occurred during a
22~GHz flare.

\par The bottom frame of Figure 1 is the 17.6~GHz (the highest useable
frequency) light curve from our more recent AMI monitoring. The 2013 flare is
on the left hand side. The outlier data points with large error bars during the
flare decay are affected by poor weather (precipitation). We were unable to
correct for the weather and they should not be considered as evidence of rapid
flux changes. The 2015 flare appears on the right hand side of the figure. The
VLBA observations (and subsequent coordinated X-ray observations) were
triggered on 2015 March 31. The flare was still rising and approaching the peak
flux density, so the timing was ideal for our program -- the maximum brightness
for study during the flare rise. Knowing that the detected fluxes were going to
be small at the crucial frequency of 43~GHz, achieving maximum brightness was
essential to the success of the observation. The other interesting feature of
the light curve is timing of the HST COS observation with the G140L grating on
2014 December 13. The flare was already beginning at this time at 17.6~GHz, and as
discussed in the next two sections, due to synchrotron self absorption (SSA)
optical depth affects the 43~GHz flare began at least a few days to weeks
earlier. Again, this is fortuitous timing of the HST observation. All three of
the far UV HST observations serendipitously occurred during times of radio
monitoring.

\begin{figure}
\begin{center}
\includegraphics[width= 0.8\textwidth ]{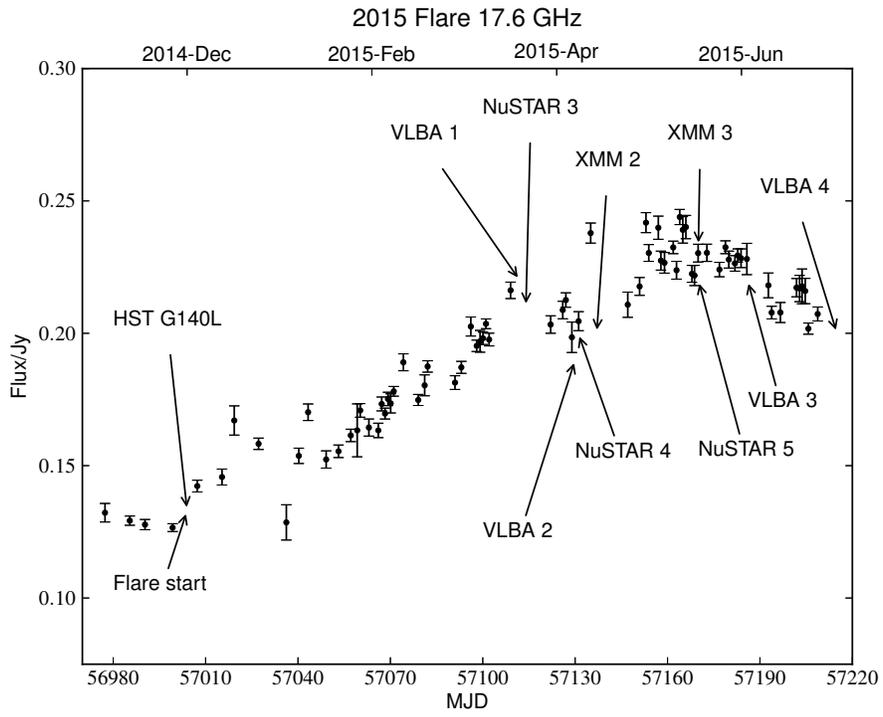}

\caption{The AMI 17.6~GHz light curve of the 2015 flare. Note the
dates of the observations discussed in this article.}
\end{center}
\end{figure}

\section{VLBA Observations}
The VLBA project, BR205, successfully acquired data from all ten
stations during each of the four epochs of observation. The data were
correlated on the VLBA correlator \citep{deller2011} and calibrated with NRAO's
Astronomical Imaging Processing System using the ParselTongue interface
\citep{kettenis2006}.

Our observations were phase-referenced to J1302+5748 ($1.3^{\circ}$
from Mrk\,231) at 8.4, 15, 22, and 43~GHz, following the strategies
described in \citet{rey09} for project BP124. J1311+5513 ($\gtrsim
80$~mJy of unresolved flux density at 43~GHz) was used as a
secondary calibrator to check the quality of the phase referencing.
For each epoch of our VLBA monitoring, two 4-hour observations with
almost continuous recording at 2~Gbps (256~MHz bandwidth, dual
polarizations) were acquired in order to achieve the SNR
\textbf{(Signal to Noise Ratio)} required to resolve the putative
sub-components of the 43~GHz core and to maximize the likelihood of
successful phase-referencing and the subsequent self-calibration of
the weak target source, Mrk\,231. For scheduling purposes, the 4
hour blocks were split into two consecutive days except for the last
epoch in which both blocks occurred in a continuous time interval.
After calibration overhead, this resulted in 2.6~hours on Mrk\,231
per epoch, that was split as 0.25, 0.35, 0.8, and 1.2 hours at
frequencies of 8, 15, 22, and 43~GHz, respectively. All 4 stokes
parameters were correlated and we were able to achieve naturally
weighted image RMS values in Stokes I of approximately 50, 100, 50,
and 70~$\mu$Jy at 8.4, 15, 22, and 43~GHz respectively.

The observations were generally successful except for one major
unexpected technical issue. Due to scheduling constraints, we could
not be granted a continuous 8~hour observation. The compromise of
splitting the observation into 4~hour increments on consecutive days
proved deleterious in epochs 2 and 3 as the two segments were
observed at almost the same LST \textbf{(Local Sidereal Time)}. The
resulting beam was poorly shaped due to insufficient $u-v$ coverage
and the resolution was inadequate for resolving structures along the
jet. We analyze the resulting data even though this circumstance
made it impossible to directly monitor motion along the jet (a
primary objective of the program).

\begin{table*}
 \centering
\caption{Summary of Model Fits to VLBA Observations} {\scriptsize
\begin{tabular}{ccccccccccccc}
 \hline
1          & 2    &     3       & 4              & 5             & 6                &  7           & 8             & 9                        &  10   & 11   & 12  \\
 Component & epoch & Flux       & Flux           & X              & X                &  Y           & Y             & Major    & Axial & PA  & Frequency \\
           &       & Density     &  Density $\sigma$ &      & $\sigma$                 &    & $\sigma$    &  Axis                  & Ratio &       &\\
           &       &   (Jy)      & (Jy)              &  (mas)         & (mas)              &   (mas)     &    (mas)        &    (mas)               &        & (deg)       &  (GHz))\\

\hline
\hline
 &    &  &  & & Epoch  &      1      &             &     &     &    &  \\
\hline
\hline
K1  &  2015.247 & 0.215 & 0.011  & 0.171 & 0.001  &0.101 & 0.001 & 0.33 &   0.83  & -80.1 & 8.42 \\
Core  &  2015.247 & 0.071 & 0.004  & 1.205 & 0.002 & 0.582 & 0.003 & 0.17 & 0.93    & 48.4 & 8.42 \\
K1  &  2015.247 & 0.093 & 0.005  & 0.186 & 0.001 & 0.103 & 0.002 & 0.29 &   0.77  & -85.5 & 15.3 \\
Core  &  2015.247 & 0.094 & 0.005  & 1.259 & 0.001 & 0.601 & 0.002 & 0.12 & 0.00    & 85.8 & 15.3 \\
K1  &  2015.247 & 0.052 & 0.004  & 0.148 & 0.002  &0.149 & 0.003 & 0.28 &   0.72  & -85.3 & 22.2\\
Core  &  2015.247 & 0.083 & 0.006  & 1.236 & 0.001 & 0.657 & 0.002 & 0.11 & 0.63    & 88.2 & 22.2 \\
Core  &  2015.247 & 0.045 & 0.005  & 1.253 & 0.001  &0.613 & 0.001 & 0.10 & 0.00  & 49.6 & 43.1\\
K1  &  2015.247 & 0.018 & 0.002  & 0.137 & 0.002  &0.097 & 0.004 & 0.29 &  0.52  & 73.5 & 43.1\\
K2  &  2015.247 & 0.002 & 0.001  & 0.939 & 0.017 & 0.449 & 0.030 & 0.00 &  1.00    & 0.00 & 43.1\\
\hline \hline
 &    &  &  & & Epoch  &      2      &             &     &     &    &  \\
\hline \hline
K1  &  2015.296 & 0.228 & 0.011  & 0.285 & 0.001  &0.095 & 0.001 & 0.45 &   0.62  & 68.3 & 8.42 \\
Core  &  2015.296 & 0.069 & 0.003  & 1.344 & 0.004 & 0.587 & 0.004 & 0.18 & 0.00    & -58.2 & 8.42 \\
K1  &  2015.296 & 0.098 & 0.005  & 0.186 & 0.002  &-0.023 & 0.002 & 0.26 &   0.88  & -83.2 & 15.3 \\
Core  &  2015.296 & 0.094 & 0.005  & 1.250 & 0.005 & 0.471 & 0.002 & 0.07 & 0.61    & -14.9 & 15.3 \\
K1  &  2015.296 & 0.057 & 0.004  & 0.225 & 0.002  &0.080 & 0.002 & 0.27 &   0.95  & 52.9 & 22.2\\
Core  &  2015.296 & 0.083 & 0.006  & 1.295 & 0.001 & 0.580 & 0.002 & 0.17 & 0.45    & 24.2 & 22.2 \\
Core  &  2015.296 & 0.044 & 0.004  & 1.315 & 0.002  &0.580 & 0.002 & 0.14 &   0.00  & 67.5 & 43.1\\
K1 &  2015.296 & 0.018 & 0.002  &    0.214 & 0.005 & 0.073 & 0.004 & 0.25 & 0.57    & 88.1 & 43.1\\
\hline \hline
 &    &  &  & & Epoch  &      3      &             &     &     &    &  \\
\hline \hline
K1  &  2015.452 & 0.259 & 0.013  & 0.186 & 0.001  & 0.007 & 0.001 & 0.43 &   0.62  & 68.4 & 8.42 \\
Core  &  2015.452 & 0.070 & 0.004  & 1.237 & 0.015 & 0.485 & 0.005 & 0.33 & 0.00    & 19.9 & 8.42 \\
K1  &  2015.452 & 0.104 & 0.005  & 0.137 & 0.003  &-0.052 & 0.002 & 0.33 &   0.57  & 26.9 & 15.3 \\
Core  &  2015.452 & 0.077 & 0.004  & 1.183 & 0.004 & 0.422 & 0.003 & 0.31 & 0.00    & 45.4 & 15.3 \\
K1  &  2015.452 & 0.057 & 0.004  & 0.110 & 0.002  & -0.083 & 0.002 & 0.24 &   0.87  & 86.4 & 22.2\\
Core  &  2015.452 & 0.055 & 0.004  & 1.175 & 0.002 & 0.401 & 0.002 & 0.14 & 0.32    & 21.3 & 22.2 \\
Core  &  2015.452 & 0.027 & 0.003  & 1.233 & 0.003  &0.408 & 0.003 & 0.08 &   0.38  & 78.7 & 43.1\\
K1 &  2015.452 & 0.020 & 0.002  &    0.150 & 0.004 & -0.092 & 0.004 & 0.24 & 0.87    & 86.4 & 43.1\\
\hline \hline
 &    &  &  & & Epoch  &      4      &             &     &     &    &  \\
\hline \hline
K1  &  2015.529 & 0.262 & 0.013  & 0.155 & 0.002  & 0.006 & 0.004 & 0.29 &   0.71  & -64.3 & 8.42 \\
Core  &  2015.529 & 0.063 & 0.003  & 1.181 & 0.010 & 0.482 & 0.015 & 0.21 & 0.00    & -8.2 & 8.42 \\
K1  &  2015.529 & 0.096 & 0.005  & 0.085 & 0.002  &-0.058 & 0.003 & 0.28 &   0.44  & -88.1 & 15.3 \\
Core  &  2015.529 & 0.048 & 0.002  & 1.146 & 0.004 & 0.426 & 0.007 & 0.18 & 0.00    & 63.0 & 15.3 \\
K1  &  2015.520 & 0.066 & 0.005  & 0.066 & 0.001  &-0.071 & 0.002 & 0.24 &   0.54  & -80.3 & 22.2\\
Core  &  2015.529 & 0.042 & 0.003  & 1.142 & 0.002 & 0.423 & 0.003 & 0.08 & 0.00    & 64.1 & 22.2 \\
Core  &  2015.529 & 0.032 & 0.003  & 1.205 & 0.002 & 0.422 & 0.004 & 0.21 & 0.00    & 67.0 & 43.1\\
K1  &  2015.529 & 0.025 & 0.002  & 0.071 & 0.003  &-0.097 & 0.005 & 0.22 &   0.69  & -69.5 & 43.1\\
K3  &  2015.529 & 0.003 & 0.001  & 1.049 & 0.027 & 0.247 & 0.042 & 0.00 & 1.00   & 0.00 & 43.1\\

\hline
\hline
\end{tabular}}

\end{table*}

\par Figure 2 is the 17.6~GHz light curve of the 2015 flare. The
flare initiates at the beginning to the middle of December 2014. The
relevant VLBA, HST, XMM and NuSTAR observations times are indicated
relative to the light curve. Table 1 lists the results of our model
fits to the VLBA observation. There are four components listed in
column (1), the core, K1: a stationary secondary 0.8 pc from the
core that has been detected in previous VLBA observations and two
new components, K2 and K3. We discuss whether K2 and K3 might be the
same component in this section. New components are only detected in
epochs 1 and 4, presumably due to an inopportune beam shape in
epochs 2 and 3. Column (2) is the date of the observation. The flux
density of the component is listed in column (3). The uncertainty in
the flux density is listed in the column (4). The next four columns are the X
(east-west) and Y (north-south) coordinates (relative to the phase centre) of
each fitted component and the associated uncertainty. Notice that the origin
(phase centre) is close to K1, not the core, due to the fact that the early
position determination was done at low frequency where the steep spectrum K1
component dominates \citep{rey09}. Columns (9) -- (11)
describe the fitted elliptical Gaussian or point source models for each
component. We list the major axis, the axial ratio and the position angle. The
last column is the frequency of observation.
\begin{figure*}
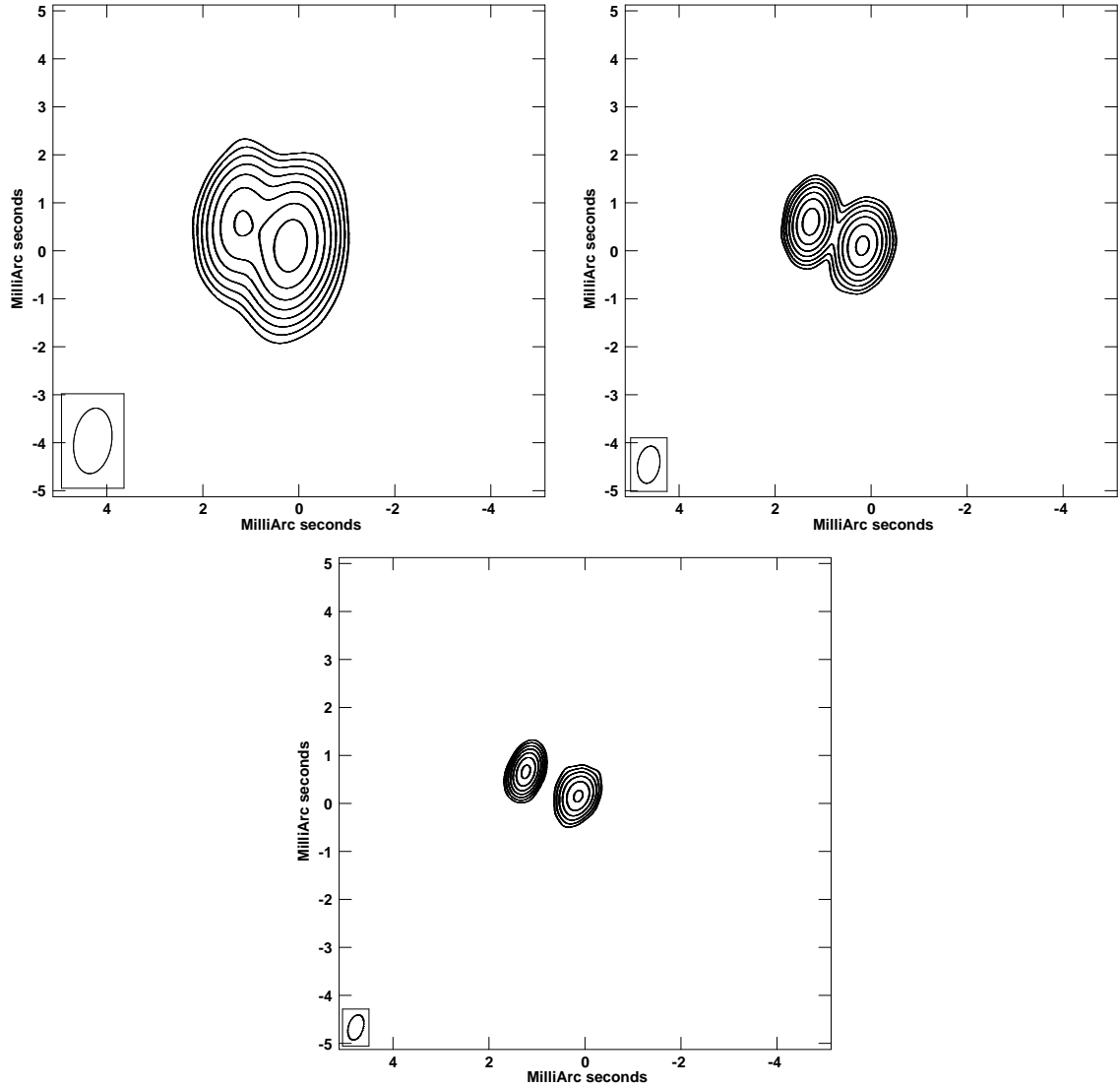

\begin{center}
\includegraphics[width= 0.45\textwidth,angle =-90]{5080_1.PS}
\includegraphics[width= 0.45\textwidth,angle =-90]{5080_2.PS}
\includegraphics[width= 0.45\textwidth,angle =-90]{5080_3.PS}

 \caption{VLBA images at 8.4 (top left), 15.3 (top right) and
22.2~GHz (bottom) from epoch 1. In each case the bottom contour is
0.1 mJy/beam increasing in factors of 2. The data have been weighted
with Briggs's weighting scheme \citep{briggs95} using a robustness
parameter of 0.}
\end{center}
\end{figure*}
\begin{figure*}
\begin{center}
\includegraphics[width= 0.45\textwidth,angle =-90]{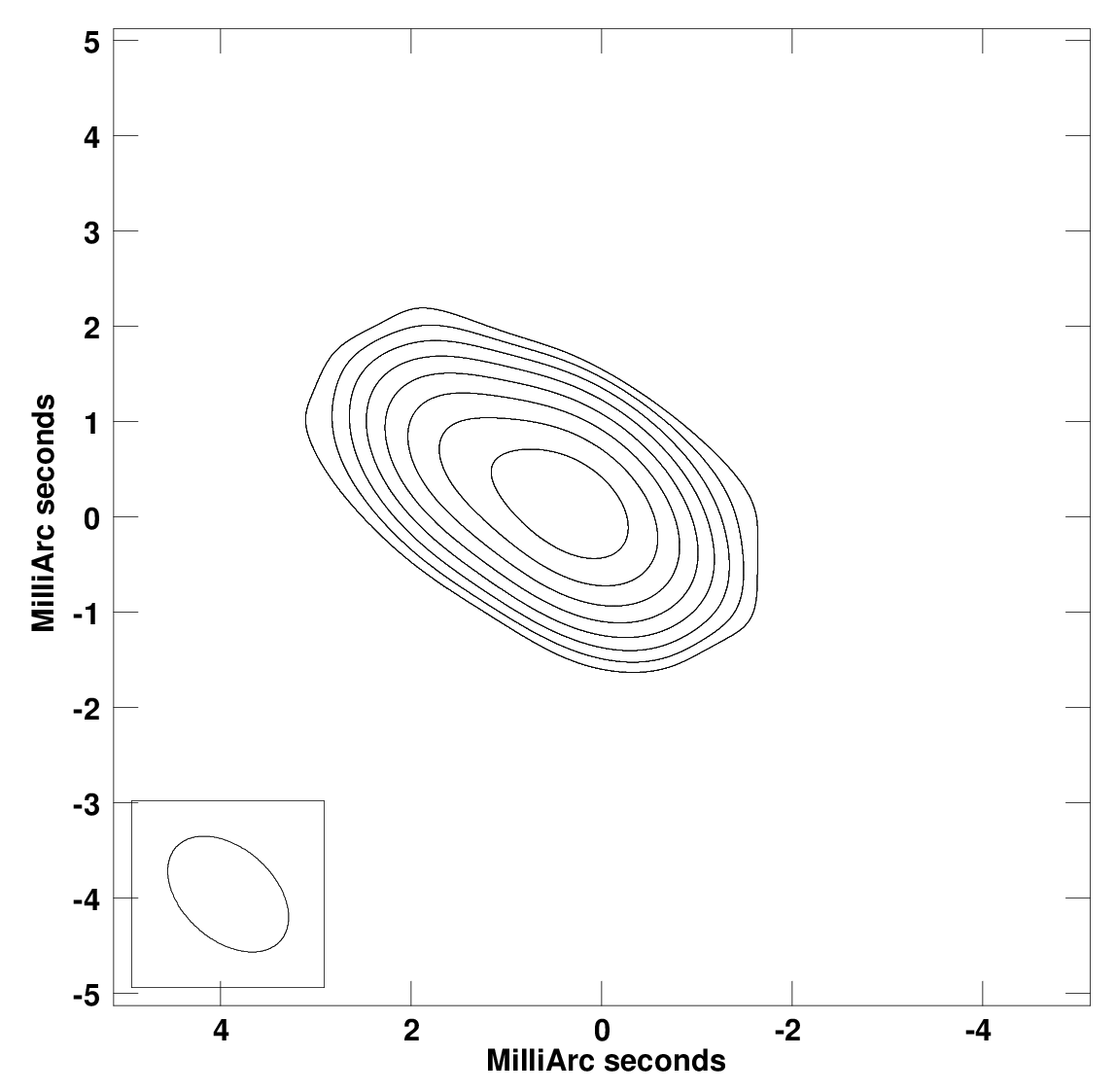}
\includegraphics[width= 0.45\textwidth,angle =-90]{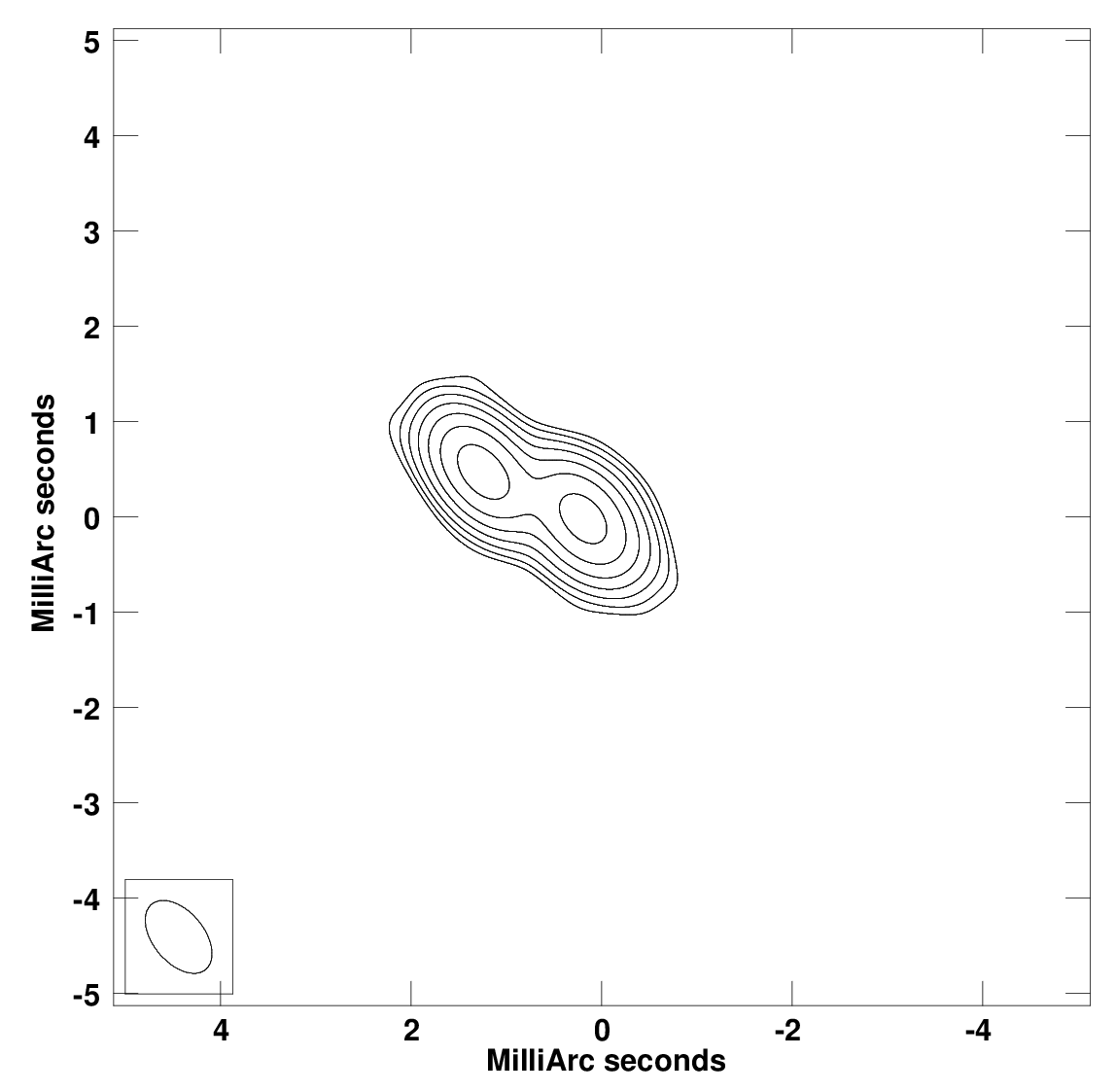}
\includegraphics[width= 0.45\textwidth,angle =-90]{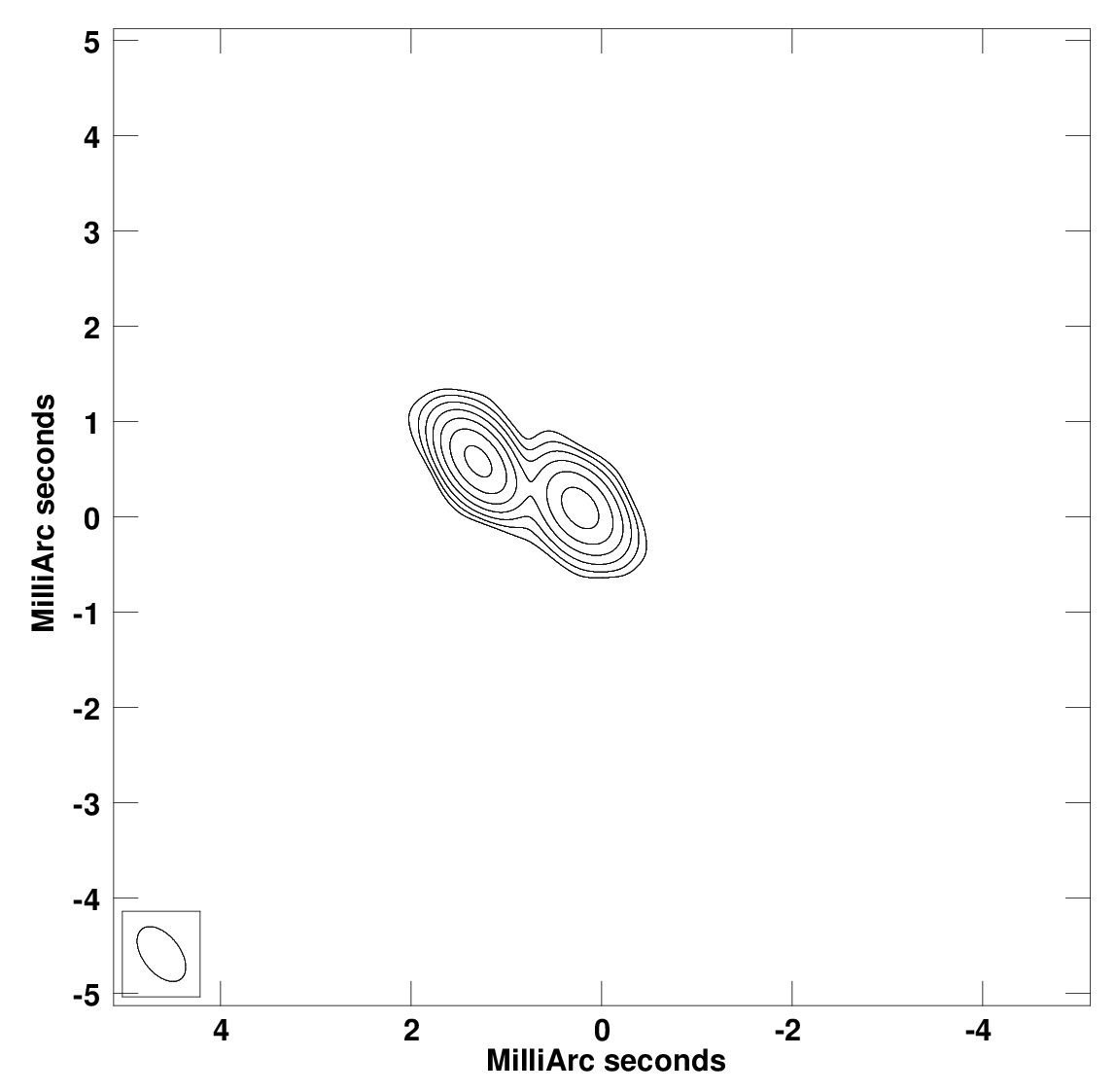}
\setcounter {figure} {2}
  \caption{ (continued) VLBA images at 8.4 (top left), 15.3 (top right) and
22.2~GHz (bottom) from epoch 2. In each case the bottom contour is
0.1 mJy/beam increasing in factors of 2. The data have been weighted
with Briggs's weighting scheme \citep{briggs95} using a robustness
parameter of 0.}
\end{center}
\end{figure*}
\begin{figure*}
\begin{center}
\includegraphics[width= 0.45\textwidth,angle =-90]{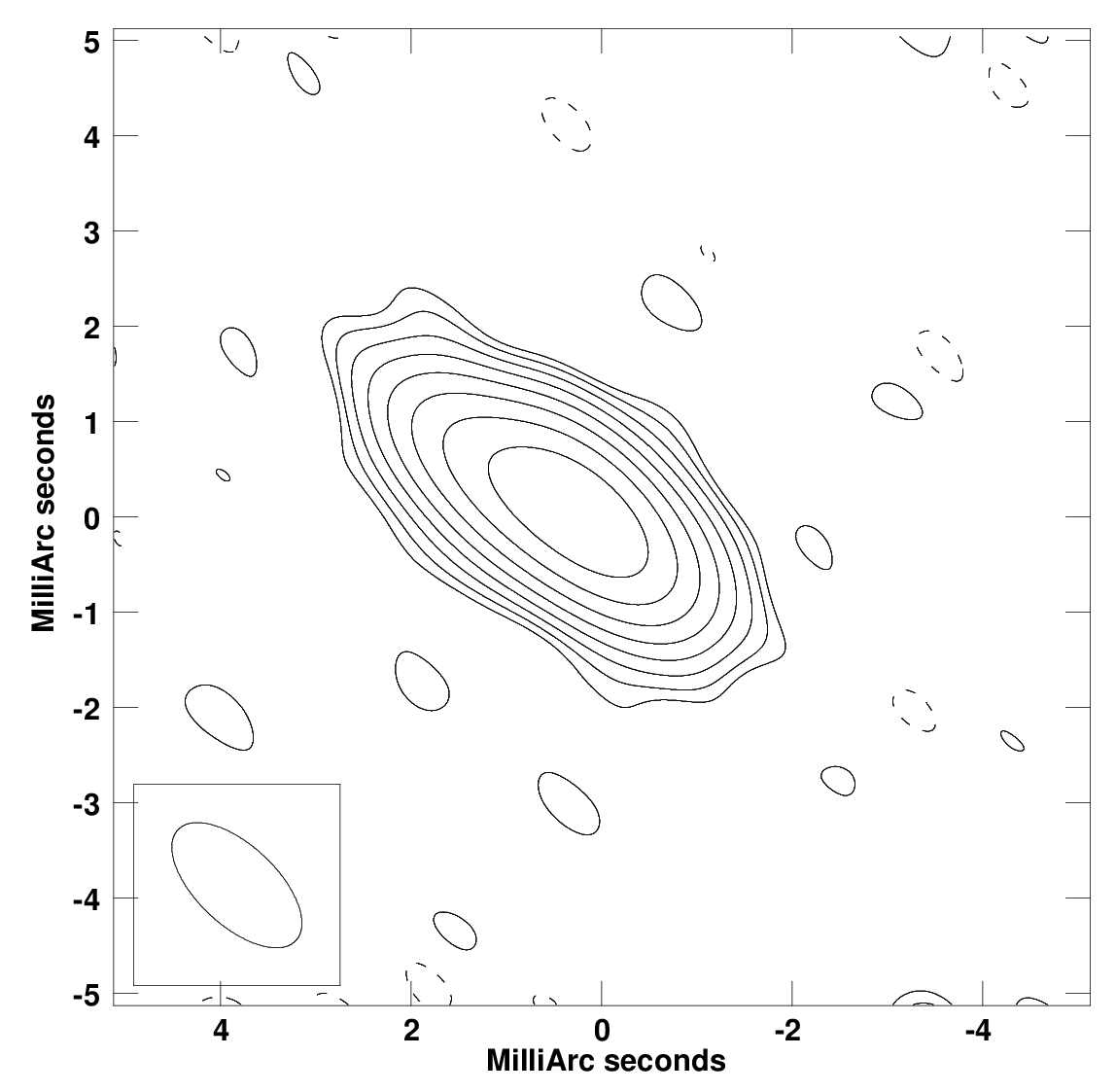}
\includegraphics[width= 0.45\textwidth,angle =-90]{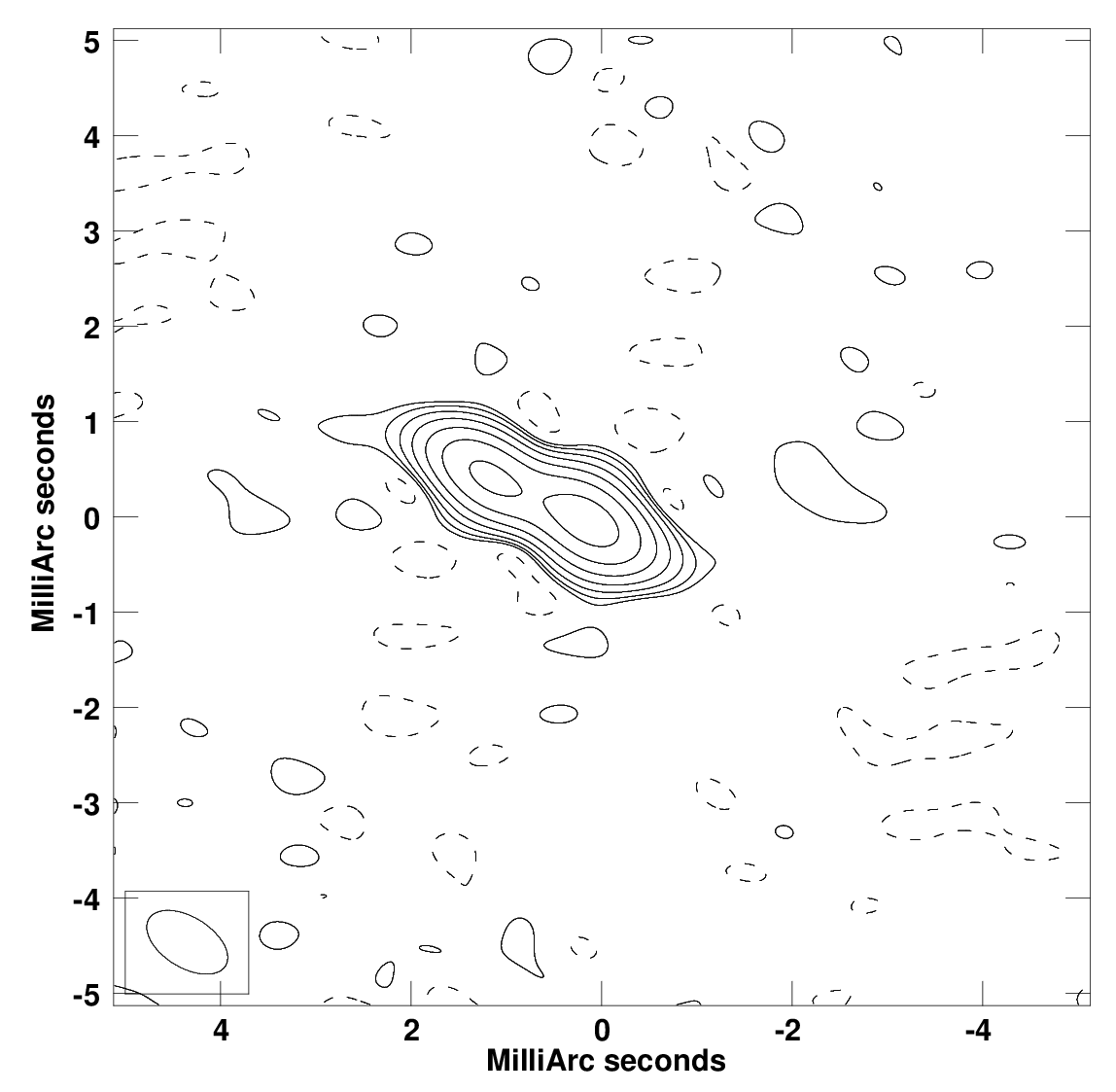}
\includegraphics[width= 0.45\textwidth,angle =-90]{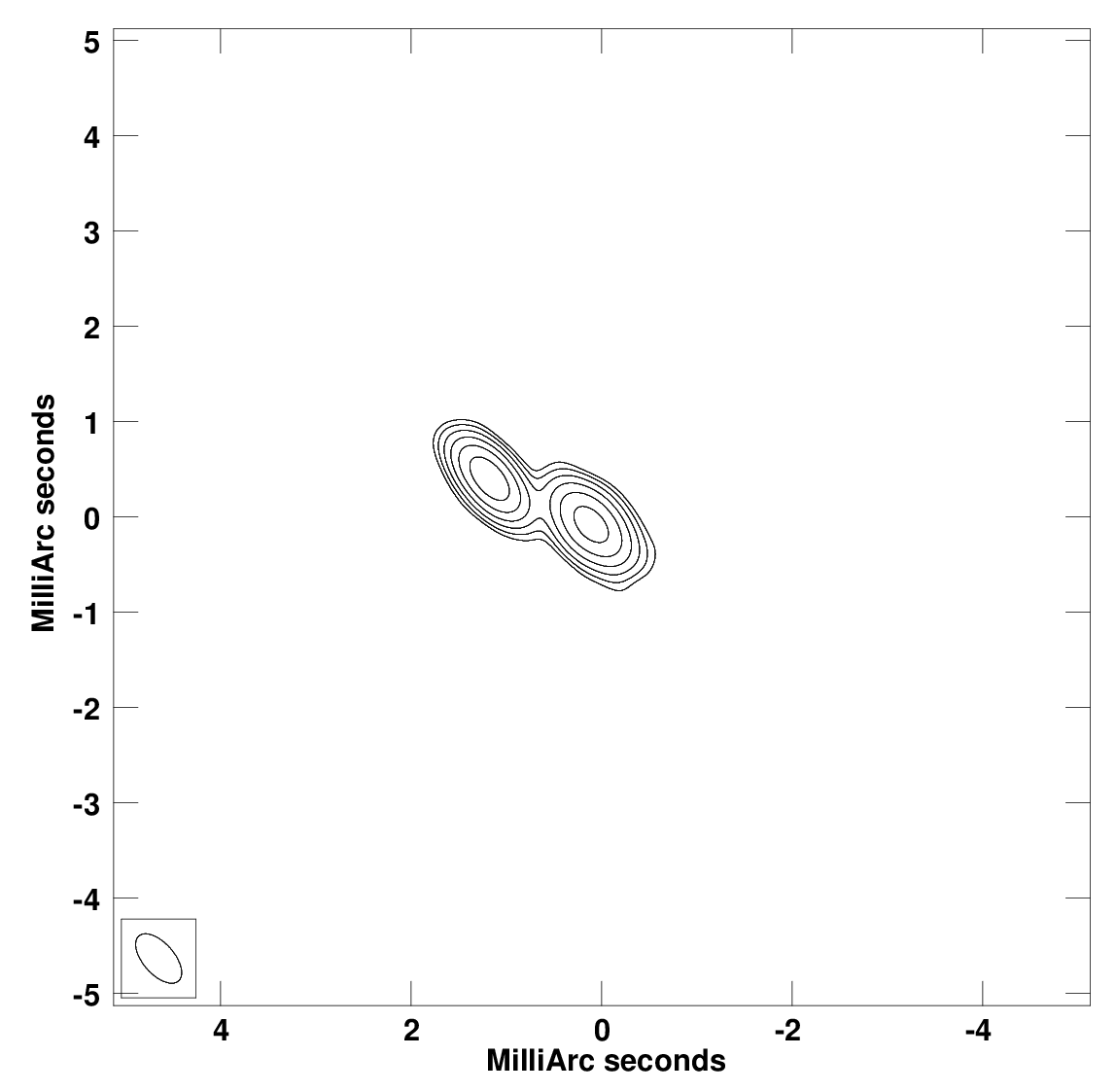}
\setcounter {figure} {2}
  \caption{(continued) VLBA images at 8.4 (top left), 15.3 (top right) and
22.2~GHz (bottom) from epoch 3. In each case the bottom contour is
0.1 mJy/beam increasing in factors of 2. The data have been weighted
with Briggs's weighting scheme \citep{briggs95} using a robustness
parameter of 0.}
\end{center}
\end{figure*}
\begin{figure*}
\begin{center}
\includegraphics[width= 0.45\textwidth,angle =-90]{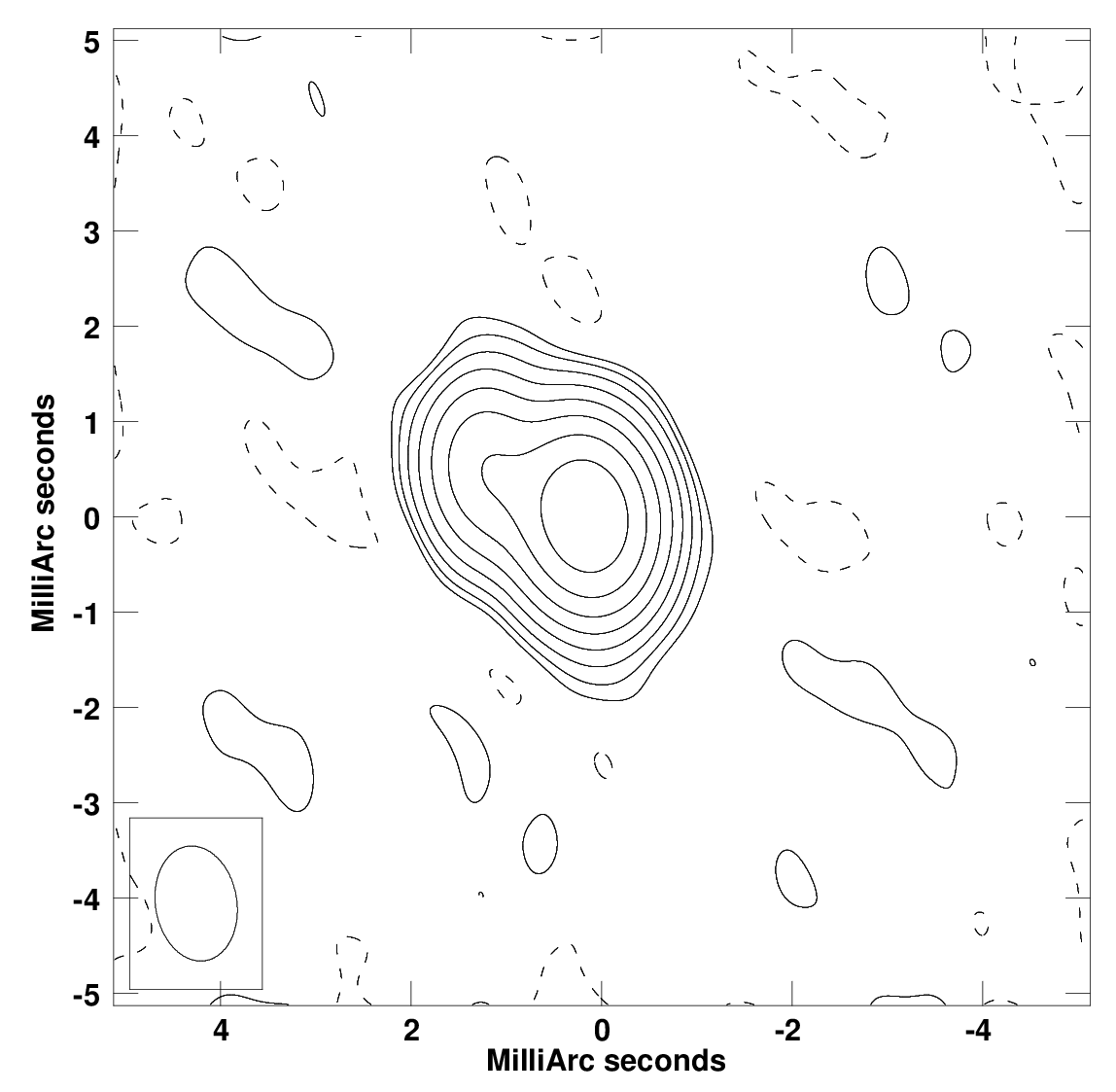}
\includegraphics[width= 0.45\textwidth,angle =-90]{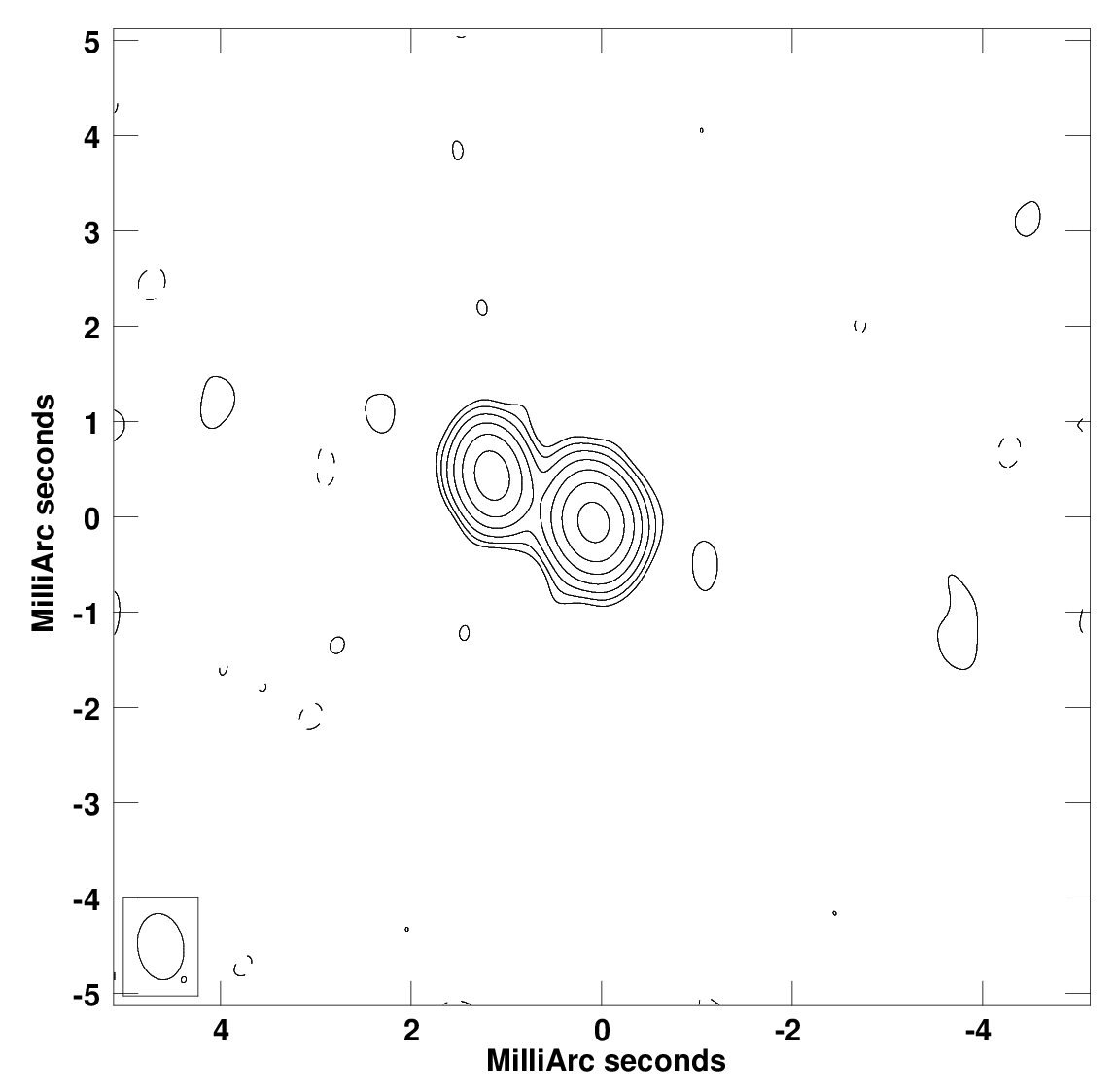}
\includegraphics[width= 0.45\textwidth,angle =-90]{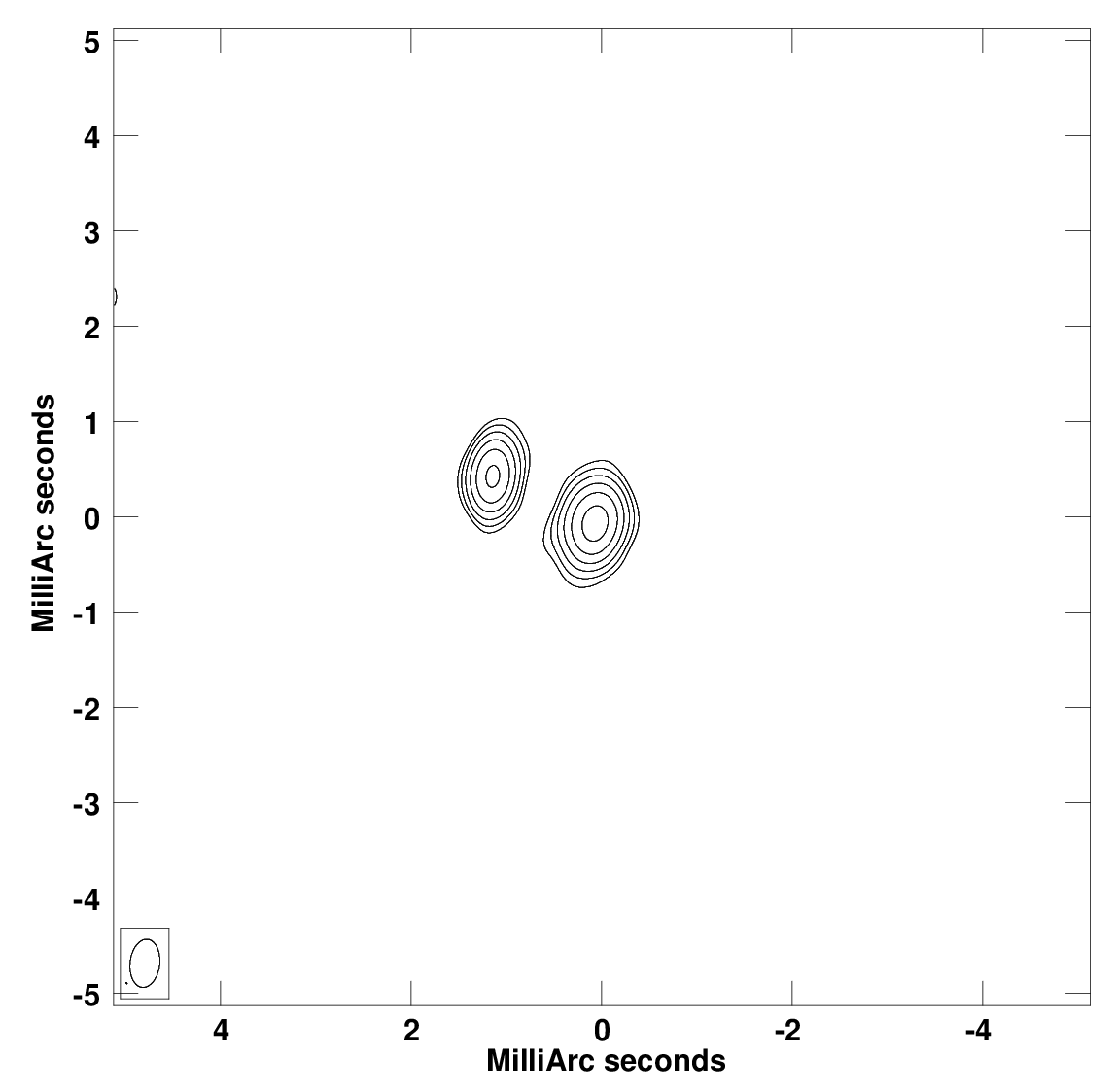}
\setcounter {figure} {2}
 \caption{(continued) VLBA images at 8.4 (top left), 15.3 (top right) and
22.2~GHz (bottom) from epoch 4. In each case the bottom contour is
0.1 mJy/beam increasing in factors of 2. The data have been weighted
with Briggs's weighting scheme \citep{briggs95} using a robustness
parameter of 0.}
\end{center}
\end{figure*}
\par In Figure 3, we show the radio images for 8.4, 15.3 and
22.2~GHz. These images are relatively low resolution for this study
and do not indicate any interesting nuclear morphology. The
resultant flux densities of the core are important for determining
the core spectrum in each epoch. These images are similar to the low
resolution images in \citet{rey09}. At 8.4~GHz, component K1 is
dominant and the source is only partially resolved. At 15.3~GHz, the
two components are resolved with roughly equal flux densities.
Finally, at 22.2~GHz, the core is brighter except during the flare
decay. The differences in apparent morphology at a given frequency are
primarily a result of beam shape. Notice the elongated beam shape
along the direction from the core to K1 in epochs 2 and 3. This
greatly degraded the resolution along the jet direction in the
critical 43.1~GHz observations during these epochs.
\begin{figure*}
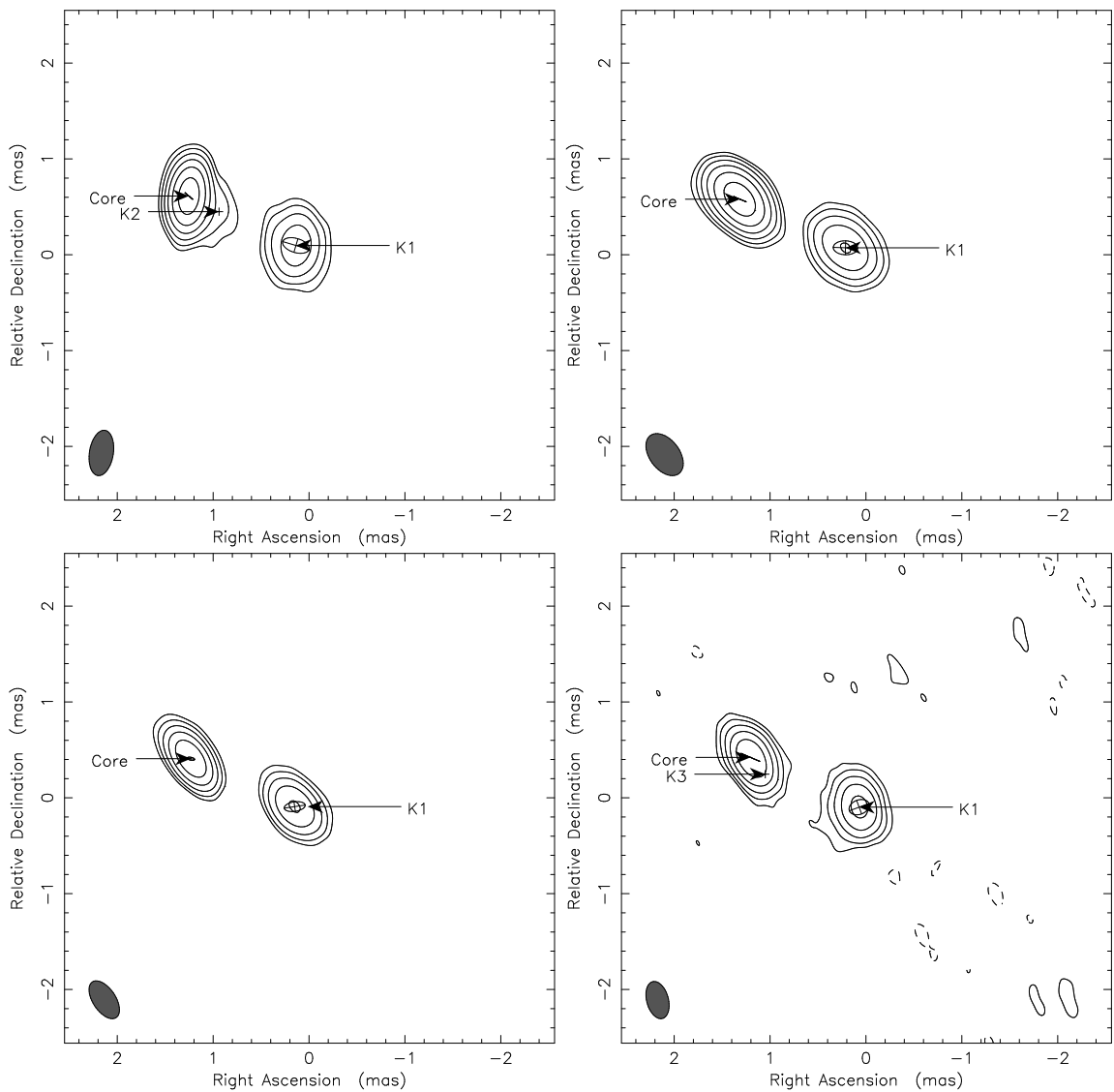

\begin{center}
\includegraphics[width= 0.45\textwidth,angle =-90,clip=true]{br205a.J1256+5652_4.ps}
\includegraphics[width= 0.45\textwidth,angle =-90,clip=true]{br205b.J1256+5652_4.ps}\\
\includegraphics[width= 0.45\textwidth,angle =-90,clip=true]{br205c.J1256+5652_4.ps}
\includegraphics[width= 0.45\textwidth,angle =-90,clip=true]{br205d.J1256+5652_4.ps}

\caption{Naturally weighted VLBA images at 43.1~GHz. The top row
from left to right are epochs 1 and 2. The bottom row from left to
right are epochs 3 and 4. The bottom contour is at 0.085 mJy/beam in
each case and contours increase in factors of 2. The crosses and
ellipses with inscribed crosses denote the locations of point source
components and the locations and dimensions of the Gaussian
elliptical components, respectively. Note that the major axis of the
elliptical Gaussian model for the core tends to point along the jet
direction. It is also very elongated in this direction during the
earlier stages of the flare.}
\end{center}
\end{figure*}
\begin{figure*}
\begin{center}
\includegraphics[width= 0.8\textwidth,angle =0]{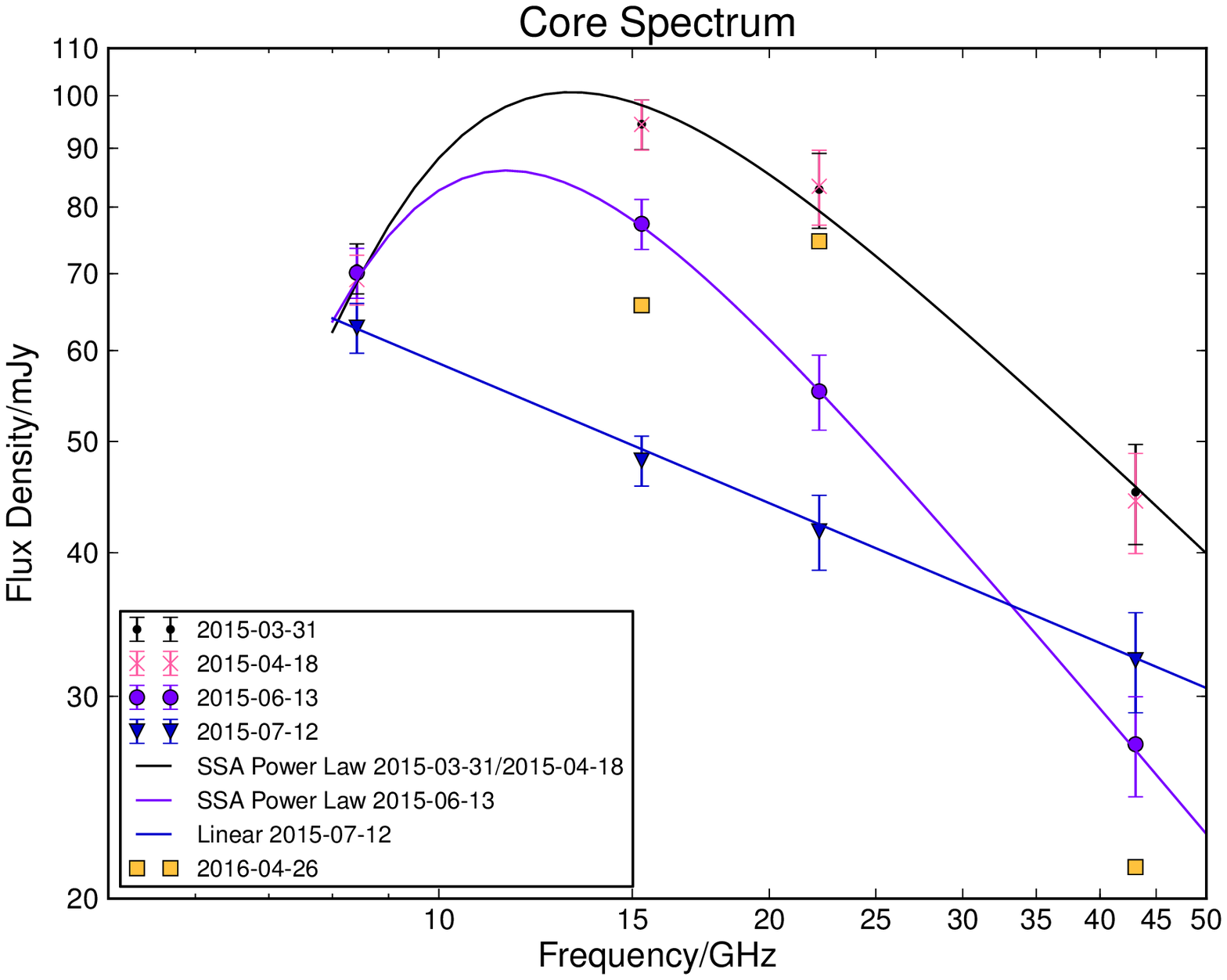}

\caption{The spectrum of the unresolved radio core in the four
epochs. The data are from Table 1. The continuous curves are SSA
power law fits to the data. We have also added the flux density of
the 2006.32 unresolved nucleus for historical reference to the flare
described in \citep{rey09}.}
\end{center}
\end{figure*}

\begin{figure*}
\begin{center}
\includegraphics[width= 0.8\textwidth,angle =0]{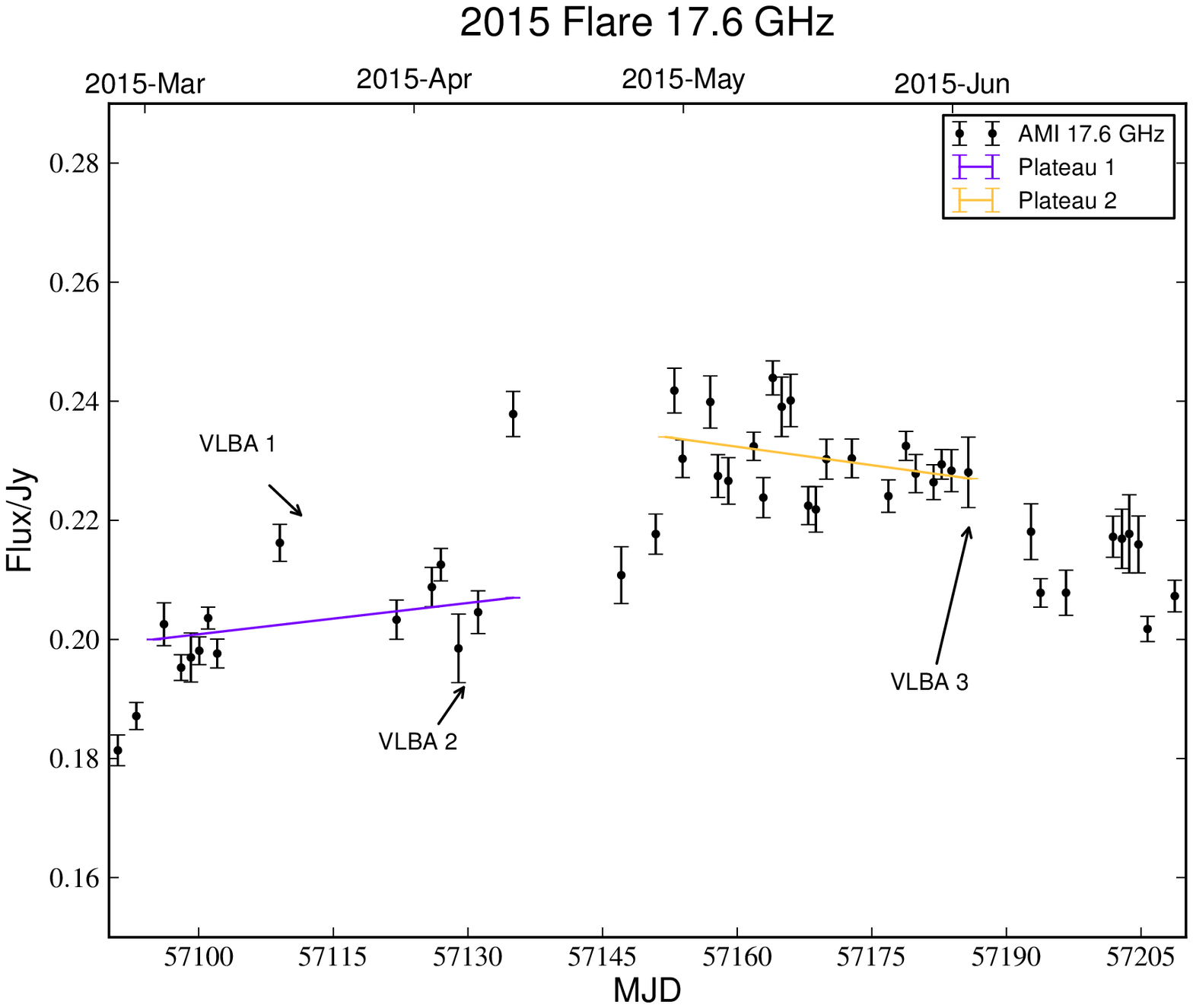}
\caption{A close up of the active plateau regions of the light
curve. These are high states associated with the peak of jet
activity.}
\end{center}
\end{figure*}
\begin{figure*}
\begin{center}
\includegraphics[width= 0.8\textwidth,angle =0]{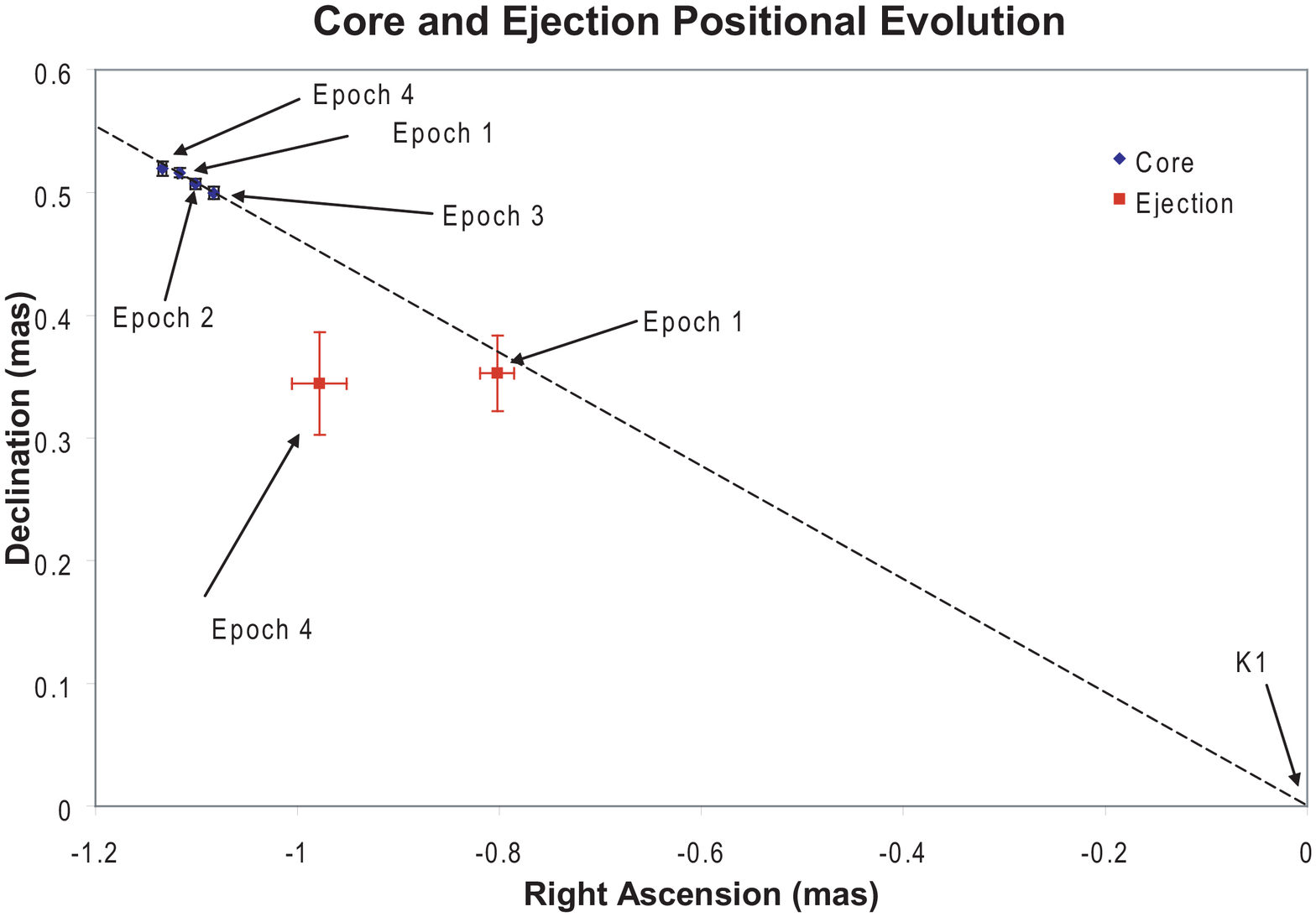}
\caption{The position of the core and the ejection relative to the
astrometric origin K1.}
\end{center}
\end{figure*}

\subsection{Nuclear Structure} Figure 4 shows the 43.1~GHz VLBA images from the four
epochs. The components K2 and K3 are the new ejections that are
partially resolved in epochs 1 and 4. Detecting these putative ejections was a
major goal of these observations. As noted earlier, the higher sensitivity
observations afforded by the new VLBA backend systems improved the ability to
phase reference and detect these ejecta compared to our 2006
observations. The new component could not be detected in epochs 2 and 3 due to
an unfortunate beam shape elongated along the jet direction: the direction that
needs the highest resolution in order to resolve new components ejected from
the core. The ejected component flux density in the two epochs is only 2--3~mJy.
We identify the partially resolved component in epochs 1 and 4 as K2 and K3,
respectively. \par Even though a large amount of information was lost due to
poor resolution along the jet direction in epochs 2 and 3, we did anticipate
the additional information that can be found in the core spectrum with a
multi-frequency observation plan. This aspect was achieved as all 10 VLBA
stations successfully observed in all epochs. The spectra of the
unresolved core in each epoch are shown in Figure 5. In \citet{rey09}, we used
the detailed nature of the spectral shape to argue that the spectrum of the
radio core was most likely a power law synchrotron spectrum that was seen
through a SSA screen as opposed to a synchrotron power law spectrum that was
seen through an opaque screen caused by free-free absorption. The spectrum in
epoch 1 is consistent with this interpretation.
\begin{figure*}
\begin{center}
\includegraphics[width= 0.8\textwidth,angle =0]{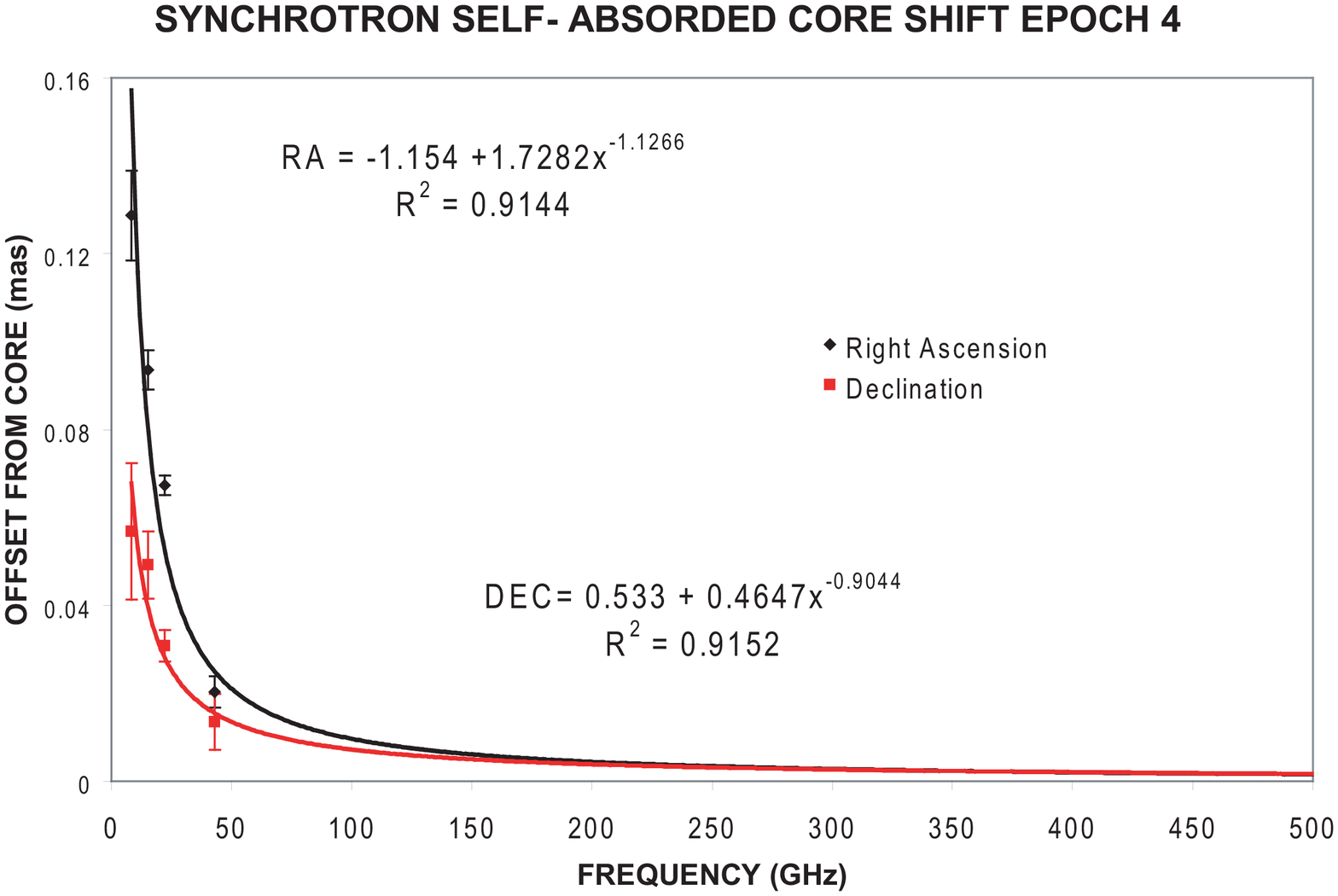}
\includegraphics[width= 0.8\textwidth,angle =0]{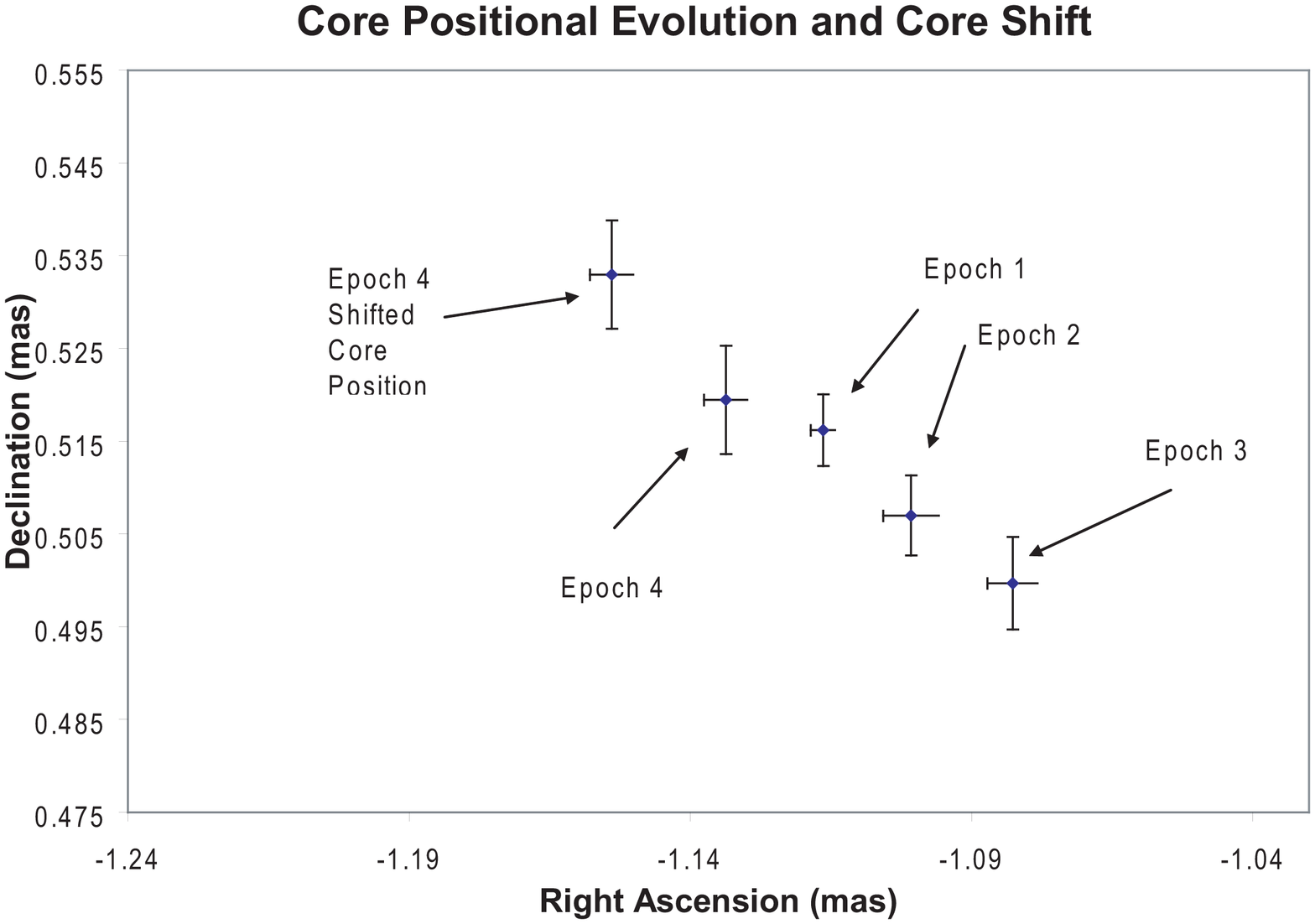}
\caption{The top frame shows the shift of the core as a function of
frequency in epoch 4 due to SSA opacity effects. The fits of the
power laws to the data are good as evidenced by the large squared
multiple regression values, $R^{2}$. The quantity, x, in the
equation for the core shift is the observed frequency in GHz. The
bottom frame shows the shifted true core location that is consistent
with zero offest at high frequency in the top frame.}
\end{center}
\end{figure*}

The plots in Figure~5 show synchrotron self-absorbed (SSA) power
law fits to the first three epochs -- a simple spectral model has
been used to describe previous observations \citep{rey09}. The first
two epochs have surprisingly similar spectra and fits. The spectral
index is $\alpha=0.91$ and the SSA optical depth is unity at 11.2~GHz (where
$\alpha$ is defined in terms of flux density as $F_{\nu}
\sim \nu^{-\alpha}$). The third epoch has $\alpha=1.13 $ and an
optical depth of unity at 10.4~GHz. The trend is ostensibly that
which would be expected as a source expands (adiabatic cooling)
and/or is synchrotron cooled or is Compton cooled by accretion disk
photons. In this scenario, the peak emission should shift to lower
and lower frequency and the spectrum should steepen at high
frequency (the highest energy electrons cool first)
\citep{mof75,tuc75,gin69}. Oddly, epoch four is a pure power law
with no turnover due to SSA absorption and $\alpha=0.41 $. There is
a fundamental physical change in the nucleus between epoch 3 and 4.
The most curious aspect of epoch 4, considering the fading of the
flare, is that the spectrum is relatively flat. It seems to
represent a disjoint epoch of low level power law emission that
supersedes the cooling plasmoid phase that occurs between epochs 2
and 3. We have included the nuclear flux densities from the 2006.32
flare that were given in \citet{rey09} for comparison purposes. In
the context of a simple SSA power law model, the nucleus was more
self-absorbed in 2006 with a steeper power law spectral index.
\par Perhaps the most unexpected finding is that the nuclear
spectrum is unchanged within the measurement accuracy of the VLBA
between epochs 1 and 2. In 18 days there is no change in the
spectrum, yet the light curve shows 20~mJy peak to peak variation at
17.6~GHz in the intervening time frame. Presumably, considering the
size of the error bars on the AMI data it is quite possible that the
intrinsic flux was actually quite stable during this time frame and
possibly even back to the middle of March 2015, with only a few mJy
variation (approximately 205~mJy -- 210~mJy) and the triggering
observation on 2016 March 27 (216~mJy) was a statistically extreme
variation arising from systematic errors. This is the only
consistent explanation of the combined AMI and VLBA data from
mid-March to mid April. The implication is there was a broad one
month plateau in the light curve that corresponded to a sustained
epoch of jet activity. With this interpretation, it is very evident
from the AMI light curve in Figure 2 that a new epoch of jet
behavior commences immediately after the epoch 2 VLBA observation,
more variable with larger flux densities, 225~mJy -- 240~mJy. The
interpretation of the peak of jet activity during the 2015 flare as
two distinct plateaus is illustrated in Figure 6. VLBA epochs 1 and
2 are in the middle and end of the plateau 1, respectively. This
could explain the similar core spectra in Figure 5. There is no
observation during plateau 2. The VLBA observation in epoch 3
appears to be during the very beginning of the decay of the more
luminous plateau 2. Unfortunately, there is no AMI data that was
available near the time of the epoch 4 observation or afterward.
\begin{figure*}
\begin{center}
\includegraphics[width= 0.8\textwidth,angle =0]{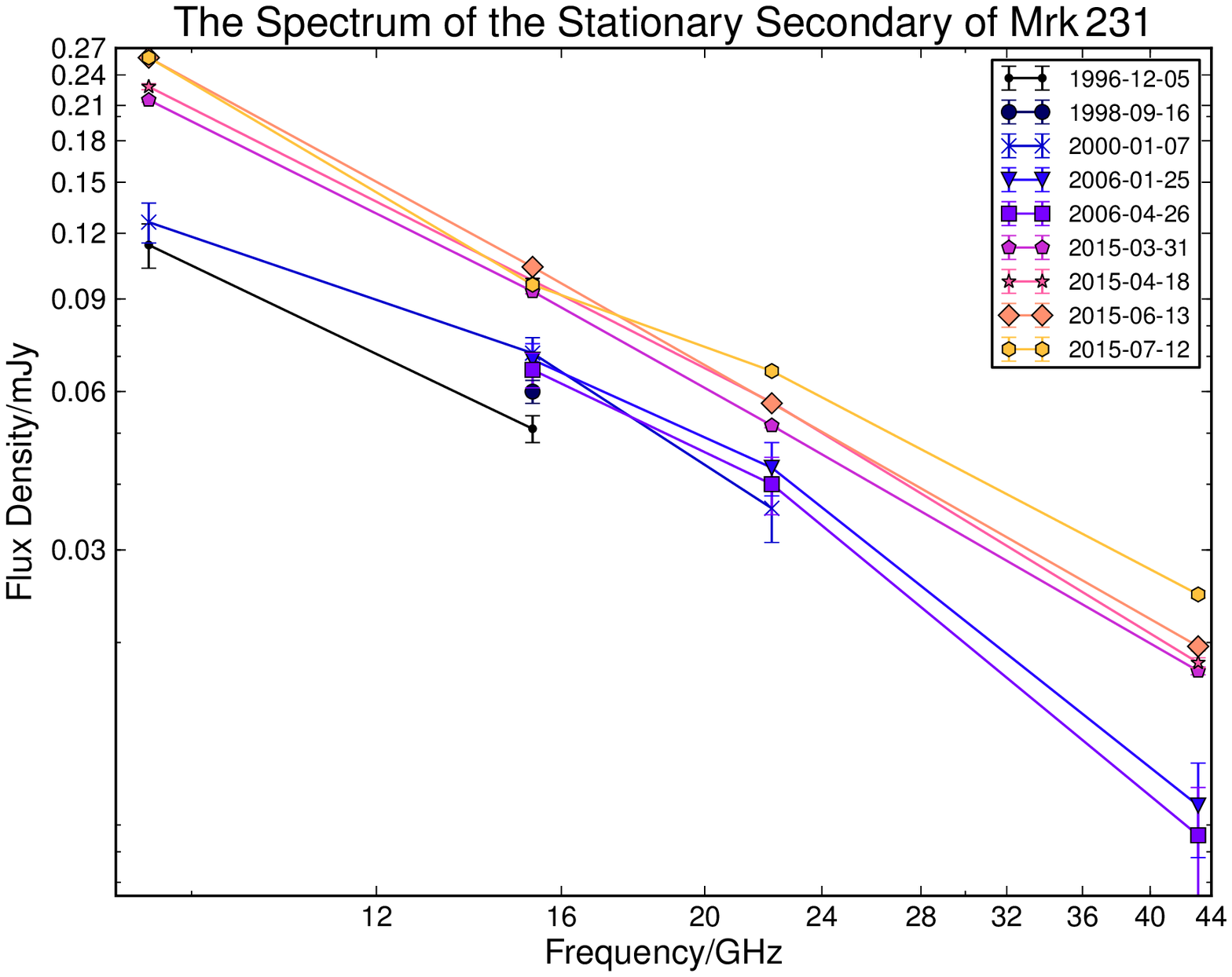}
\caption{The historic spectrum of the stationary component K1.}
\end{center}
\end{figure*}

\subsection{Ejected Components} An important diagnostic of the jet activity
near the source is provided by component motion. We can improve on the inherent
astrometric accuracy provided by our phase-referenced observations by using the
position of component K1 as a reference point. This is a very stationary
feature. In the two epochs observed in \citet{rey09}, 2006.07 and 2006.32, the
distance from the core to K1 at 43~GHz was 1.222 mas and 1.216 mas. In the four
epochs observed in this program, the separations at 43.1~GHz, were 1.230 mas,
1.212 mas, 1.192 mas and 1.247 mas in temporal order, with an average of 1.220
mas. This equates to $< 0.05$ mas of motion in 9 years. Thus, with excellent
precision we consider K1 to be a fixed origin for our astrometric coordinates.
We will explain the fluctuations in this separation during our four
observations in the following discussion. They are likely related to beam shape
and nuclear activity.

\par In contrast, the core is not a reliable origin for determining
component motion. This is not due to SSA opacity affects as are
traditionally considered in homogeneous jet models \citep{bla79}.
The spectrum is indicative of optically thin plasma at 43.1~GHz in
epochs 1 -- 3. It is likely that resolution as a consequence of beam
shape and size will determine how much of the optically thin
emission is represented in the 43.1~GHz core. In this context, it is
important to recognize that there are likely multiple discrete
ejections that are unresolved by the VLBA at 43.1~GHz. We can infer
this from two pieces of evidence. Firstly there are the abrupt jumps
in 17.6~GHz flux density that occur in the AMI light curve in Figure
2. Secondly, the Gaussian elliptical fits to the 43.1~GHz core in
Table 1 are characterized by a major axis that is very elongated
with a major axis more than half a beam width. The major axis is
approximately parallel to the direction to the stationary component,
K1. Considering the detection of two ejected components K2 and K3
and the abrupt jumps in the light curve, the existence of multiple
components along the jet direction are a natural explanation of the
43.1~GHz Gaussian fits to the unresolved core that are highly
elongated along the jet direction. This interpretation of the
nuclear fits is consistent with the core position shift that is
evident in Figure 7 from epoch to epoch. Depending on the
distribution and brightness of these ejecta, the centroid of the
unresolved beam will move. Consider the motion of the 43.1~GHz radio
core in Figure 7. In this plot, all of the coordinates are based on
making the unresolved, stationary component K1, the true origin for
astrometry. Notice that the core moves along the jet towards K1 at
epoch 2. The reason is likely two fold. First there are possible new
unresolved ejecta that might be manifested as modulations of the
light curve in Figures 2 and 6. Secondly, the elongation of the beam
shape along the jet direction inhibits the resolution of new and
older weak components that have been ejected (i.e., component K2
from epoch 1 cannot be resolved by this beam). This inability to
resolve ejecta that are directed toward K1 will result in a shift of
the centroid of the core emission toward K1. The same effect occurs
in epoch 3 as the poor resolution along the jet direction is unable
to resolve the relatively strong new ejection, K3. This is likely
the cause of the shift of the core position to its closest point
toward K1. As the core becomes less active during the fade of the
flare after epoch 3, the core drifts away from K1 as the ejecta are
fewer and weaker and therefore contribute less to the unresolved
base of the jet. The motion between epochs 3 and 4 is exaggerated by
the elongated beam along the jet direction in epoch 3. The smaller
beam along the jet direction in epoch 4 resolves out another
component which we call K3. Thus, we consider the core position in
epoch 4 our best estimate of the active galactic nucleus (AGN)
because it is based on high resolution along the jet direction and
the observation is during a state of decline of the 17.6~GHz radio
flare and there is less likely to be unresolved, strong, new
components.
\par In the absence of sufficient resolution to track component
motion in epochs 2 and 3, we use the location of the core in the
relatively quiescent epoch 4 (the AGN location) in order to
constrain the apparent velocity of the components. We note that the
shift of the core in epoch 4 to the position depicted in Figure 7
might be conservative. The jet spectrum in epoch 4 is flat spectrum,
$\alpha=0.41$. Thus, the jet might be optically thick. Therefore,
due to SSA opacity, the true core position would be slightly
upstream (to the left in Figure 7) from the epoch 4 core position.
Extrapolating the shift by fitting the SSA core shift over frequency
is logistically difficult. However, one can use the stationary
secondary position as a nearby reference point in order to register
the location of the core as a function of frequency in epoch 4. The
first attempt did not work, because the 22.2~GHz core position was
very far from the power law fit to the other 3 points. We attribute
this to the ejected component K3 that is unresolved in the 22.2~GHz
beam. Assuming a steep power law and a large SSA opacity at
15.3~GHz, we associated 5~mJy at the position of K3 with the
unresolved core at 22.2~GHz and $\approx 0$~mJy in the unresolved
core at 15.3~GHz and 8.4~GHz. We re-register the 22.2 GHz core
position based on extricating the unresolved component, K3. The
resultant powerlaw trend is plotted in the top frame of Figure 8 and
extrapolated to high frequency. It is a plot that fits a core
position in declination and right ascension with a power law for the
associated core shift from this position as a function of frequency.
This is an empirical process and the expectation is that the shift
from the true core position is zero at high frequency where the SSA
opacity is very small. The results are plotted in Figure 8 and
indicates a 0.03 mas shift of the true core position based on SSA
opacity effects. The top frame is the fit to the core shift and the
bottom frame is a closeup of the core evolution and core shift
resulting from this SSA opacity analysis. The SSA shifted core
position in epoch 4 will be denoted as the AGN location.
\par Based on the shifted core position in Figure 8, we can estimate
the velocity of the component K2. We know that component K2 was
ejected after the flare start (after 2014 December 1, see Figures 1
and 2). It moved $0.40 \pm 0.04$ mas from the AGN in $<120$ days.
Using a conversion of 1mas =0.8pc, this equates to an apparent
velocity $\beta_{app} =v_{app}/c > 3.15 \pm 0.3$ for K2. This is a
lower bound because there were likely numerous other mini-flares
(associated with local maxima) during the rise of the light curve as
discussed above. It is probably more likely that one of these
mini-flares during the flare rise is associated with the ejection of
the detected component K2. The 17.6~GHz light curve seems to
indicate that the 2015 radio flare in Mrk\,231 was composed of
multiple small superluminal ejections.
\par The component K3 is located $0.26 \pm 0.04$ mas from the
AGN. However, due to the poor beam shape in epochs 2 and 3, we have
no reliable means of constraining the velocity of K3. Finally, we
note that the trajectories for K2 (PA = -$117^{\circ}$) and K3 (PA =
-$137^{\circ}$) appear to be different. K2 appears to be on a
trajectory headed towards the steep spectrum secondary, K1.
Component K3 curiously lies below a common line in Figure 7 that
connects all the core positions and the discrete components, K1 and
K2. Reprocessing the data multiple times consistently found that K3
lies to the south of this line.

\subsection{The Stationary Secondary} If we assign the AGN location to the core at epoch 4 then as discussed above, the component
K1 moved $<0.05$ mas in 9 years which is consistent with zero
velocity with our level of measurement accuracy. However, if we
converted this to a velocity it would equal $4200$ km/s which is too
small for us to measure accurately. In \citet{rey09}, we noted that
matched (to 5~GHz) resolution VLBA observations indicated that the
secondary is seen through a shroud of free-free absorbing gas with
an emission measure of $\approx 10^{8} \mathrm{cm}^{-6}\mathrm{pc}$.
It was concluded that the steep spectrum secondary seems to be a
radio lobe associated with the jet advancing into a dense medium
(the jet is confined by ram pressure) that is also the source of the
free-free absorption. It was estimated that the long term time
averaged kinetic energy flux deposited in the stationary secondary
is $\overline{Q}\approx 10^{42}$~ergs~$\rm{s}^{-1}$.
\par Figure 9 compares the spectrum of
component K1 in our four epochs of observation to the historical
spectra that were presented in \citet{rey09}. The flux has
approximately doubled in the last 9 years. This is consistent with
the interpretation of K1 that it is a radio lobe confined by a dense
medium. The light curve in Figures 1 and 2 indicates increased radio
activity since 2011. The cumulative effect is to energize the radio
lobe with more high energy plasma. There even appears to be an
additional $> 20\%$ increase in the 8.4~GHz flux density during our
observing program between epochs 1 and 3 according to Table 1. This
is apparently real since the secondary calibrator J1311+5513 does
not show the same increase. If we associate the increase to plasma
deposited in the lobe by the strong 2013 flare then according to
Figure 1 the time of flight from the nucleus to K1 at epoch 3 is
$1.7 \,- \, 2.5$ years. However, the flux increase at 8.4~GHz is
basically a linear rise from epoch 1 to epoch 3 (see Table 1). Thus,
the time elapsed between the 2013 flare and epoch 1 is $\approx 1.5
\,- \, 2.3$ years. This time frame covers the rise of the 2013 flare
to the small flare during the light curve decay of the 2013 flare.
The peak of the 2013 flare was 1.9--2.0 years before the epoch 1
observation. The projected distance on the sky plane between the
core and K1 is 3.2 lt-yr. The apparent speed that the energy flux
propagated with is (3.2lt-yr/2.3 yr)$<\beta_{app}<$(3.2 lt-yr/1.5
yr), or $1.4<\beta_{app}< 2.1$, consistent with mildly relativistic
motion. In this scenario, based on the estimate of the velocity of
the component, K2, in the last section this value decreases as the
jet propagates toward K1. This can occur if the line of sight to the
jet increased dramatically or there was deceleration of the flow. A
change in the line of sight is inconsistent with the tight alignment
of components and the core shifts with the direction from the core
to K1 in Figure 7. This favors deceleration, which was suggested to
occur by entrainment in the dense nuclear environment of Mrk\,231
\citep{rey09}. Alternatively, if the increase of flux in K1 is from
the 2015 flare, $\beta_{app}> 9$. Neither scenario is ruled out by
the data. Using the K1 radio lobe as a crude bolometer indicates
that the jet in Mrk\,231 is becoming more energetic. Mrk\,231 might
be transforming from a radio quiet quasar to a radio loud quasar.
\begin{figure*}
\begin{center}
\includegraphics[width= 0.75\textwidth,angle =0]{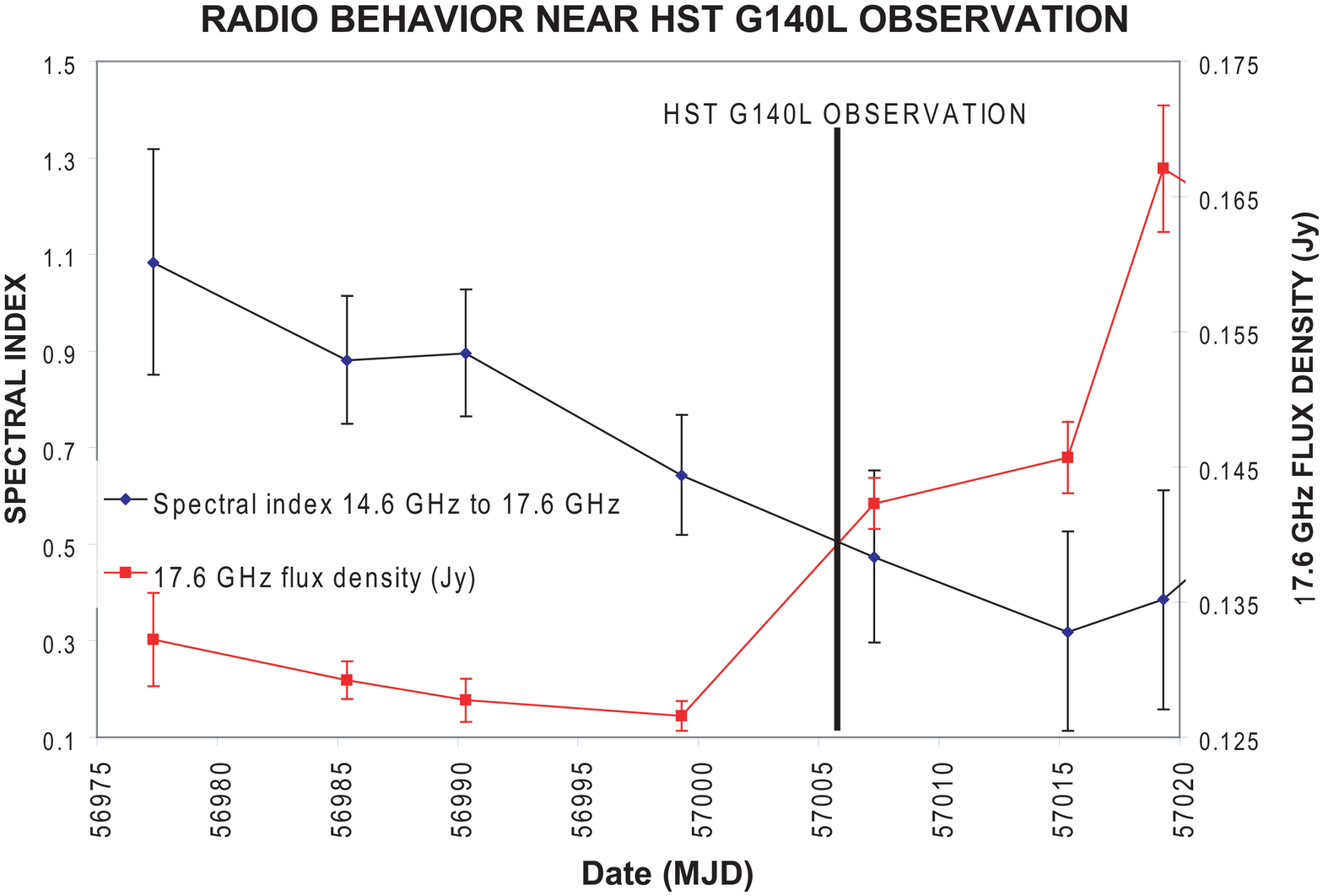}
\includegraphics[width= 0.75\textwidth,angle =0]{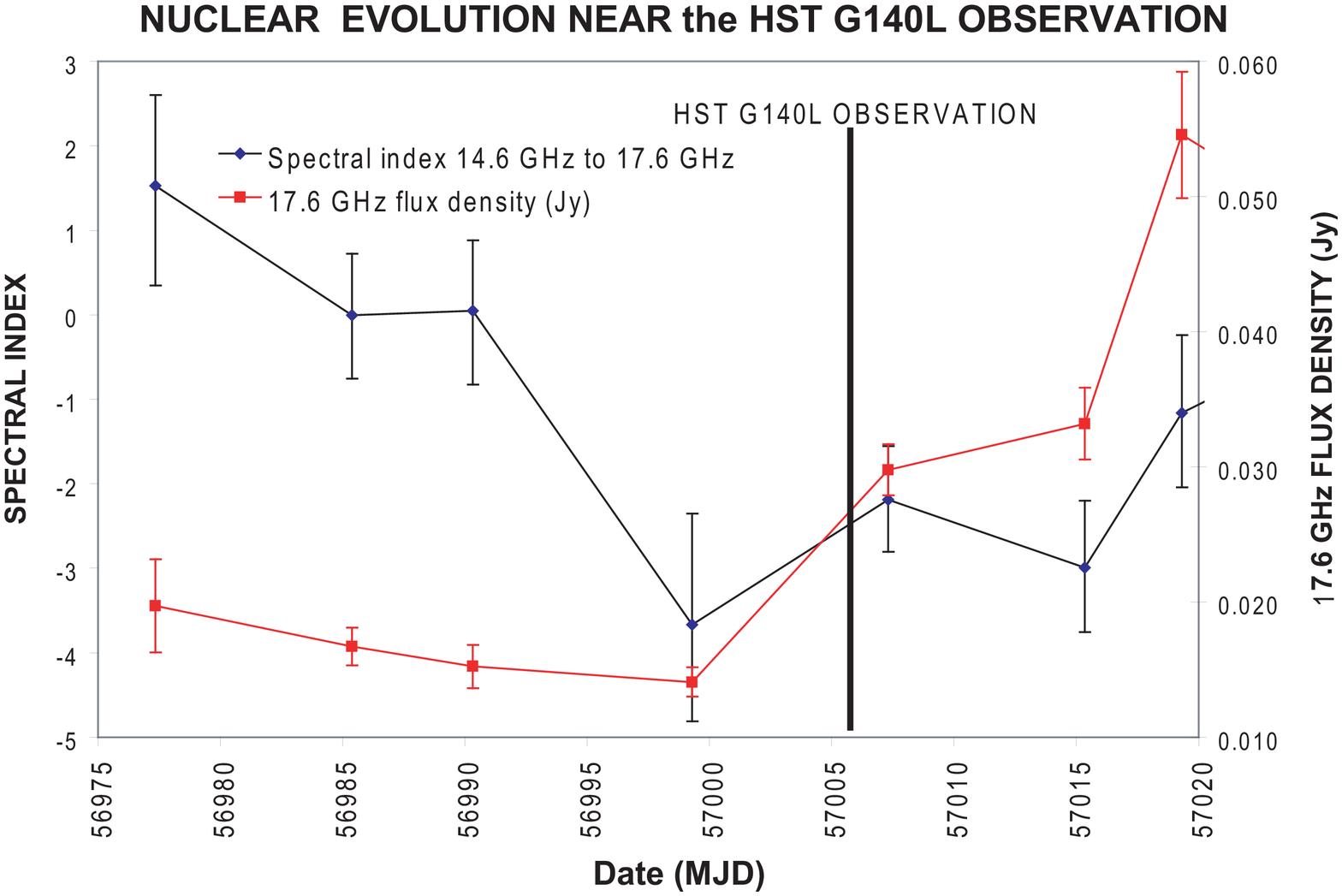}
\caption{The beginning of the major radio flare that peaked in the
spring of 2015. The top frame plots the total flux density from the
AMI data. The spectral index has a large uncertainty. However, there
is a consistent trend of spectral hardening that is evident more
than a week before the 17.6~GHz flux density begins to increase and
for 12 days afterwards. The bottom frame describes the implied
change to the nuclear component using Equations (1) -- (4) in the
text. The spectrum hardens drastically a few days days to 2 weeks
before the G140L observation. The implication is that the flux was
increasing at high frequency during this time period. As the
disturbance propagates away from the nucleus, the SSA opacity
decreases. Around the time of the G140L observations, the spectrum
begins to soften and the 17.6~GHz flux density starts to increase
rapidly: increasing by a factor of 4 in less than three weeks.}
\end{center}
\end{figure*}
\subsection{Summary of Sub-Parsec Scale Radio Structure}
In this section we used four epochs of VLBA observations in order to
constrain the morphology and time evolution of the sub-parsec scale
nuclear region during the major radio flare of 2015. The first
result, in Section 3.1, was the detection of partially resolved
substructure at 43~GHz in epochs 1 and 4. We interpret these as
ejecta associated with the major flare. The lack of the detection of
substructure in epochs 2 and 3 is likely a consequence of a poorly
shaped beam that reduced the resolution along the jet direction.
This hindered our efforts to follow the time evolution of ejected
components. In the absence of these data, we relied on determining a
precise location of the actual AGN in order to constrain the jet
speed in Section 3.2. This required a careful analysis of all four
datasets. We determined a lower bound on the apparent speed of the
ejection that was detected in epoch 1, $> 3.15$c. In Section 3.3, we
showed that the stationary secondary (the putative radio lobe)
doubled in 8.4~GHz flux density over 9 years. We even detected a
flux increase during this observing program. By associating this
flux increase with energy from the 2013 radio flare, we estimate an
average apparent jet speed of $1.4<\beta_{app}< 2.1$. Alternatively,
if the increase of flux in the stationary secondary is from the
present flare, $\beta_{app}> 9$. Neither scenario is ruled out by
the data.
\begin{figure*}
\begin{center}
\includegraphics[width= 0.8\textwidth,angle =0]{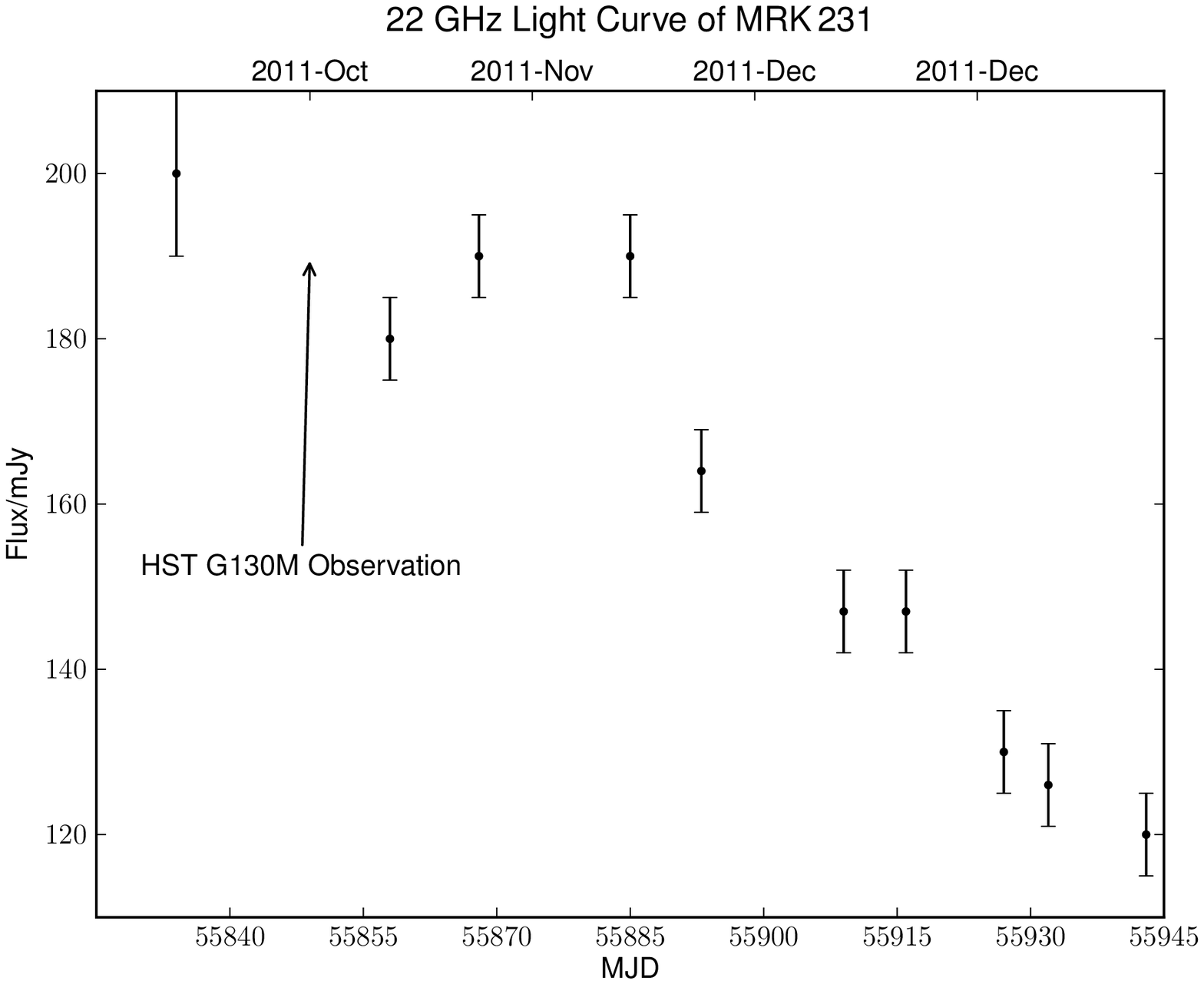}
\caption{The HST COS G130M observation during the 2011 radio flare.}
\end{center}
\end{figure*}
\section{The Far UV Jet Connection} In this section, we explore the
possible relationships between the far UV spectrum and the radio
state of the nucleus.
\subsection{The Timing of the Flare Initiation and the HST COS G140L
Observation}
The 2014 December 13 HST COS G140L observation was less than 36
hours before the first detected onset of flare activity on 2014 December 15 as
evidenced by the 17.6~GHz light curve in Figures 1 and 2. From 2014 November 15
 until 2014 December 7 the 17.6~GHz flux density was in a slow monotonic
decline from 132~mJy to 127~mJy. The next data point (MJD 57007) was 8 days
later and the flux density increased to 142~mJy. This increase was not a
statistical fluctuation or systematic observational error since the light curve
showed steady growth afterwards (except for one outlier point in bad weather 29
days later). Furthermore, the spectral index from 14.6~GHz to 17.6~GHz showed a
linear decrease (spectral flattening) from MJD 56985 until MJD 57019, going
from $\alpha=1.1$ to $\alpha=0.3$ (see the top frame of Figure 10). The
spectral index has a large uncertainty and using 14.6~GHz for our low frequency
instead of 13.8~GHz is much more reliable with AMI data. However, there is a
consistent trend of spectral hardening that is evident more than a week before
the 17.6~GHz flux density begins to increase and for 12 days afterwards.
\par The changes to the nucleus would have been
profound because the preponderance of the 17.6~GHz flux is from the
stationary secondary and the surrounding large scale radio emission
\citep{ulv99}. We can estimate these two background components. In
order to determine the flux density of the stationary secondary,
$F_{\nu}^{SS}$, we note the increase of the flux density during the
first three epochs of our VLBA program that was discussed in the
last section. This increase was much larger at 8.4~GHz than at 15.3~GHz. Thus,
we expect a different time dependence for different
frequencies. We estimate the flux densities at 14.6~GHz and 17.6~GHz
in the four VLBA epochs by extrapolating the 15.3~GHz flux density
using the two point spectral indices to 8.4~GHz and 22.2~GHz. We
find at 14.6~GHz that $F_{\nu}^{SS}(\nu =14.6 \,\rm{GHz}) =$ 99~mJy,
104~mJy and 111~mJy in epochs 1, 2 and 3, respectively. Similarly,
we find at 17.6~GHz that $F_{\nu}^{SS}(\nu =17.6) \,\rm{GHz} =$ 74~mJy, 80~mJy
and 83~mJy in epochs 1, 2 and 3, respectively. We fit
these data by powerlaws in time and extrapolate back to the time of
the HST G140L observation
\begin{equation}
F_{\nu}^{SS}(\nu =14.6 \,\rm{GHz}) = 85 \, mJy \; , \;
F_{\nu}^{SS}(\nu =17.6 \,\rm{GHz}) = 67\, mJy \; \rm {MJD\, 57005}
\;.
\end{equation}
We can find the background large scale flux density, $F_{\nu}^{LS}$,
using the simultaneous VLA and VLBA observations in December 1996
\citep{ulv99}. Being large scale in nature ($\sim 100$ pc), these
fluxes should be steady over 20 years. The VLBA images only the core
and the stationary secondary with combined flux densities of 132~mJy
and 63~mJy at 8.4~GHz and 15.4~GHz, respectively. The VLA is sensitive to
larger scale structures and detected flux densities of 203~mJy and 112~mJy at
8.4~GHz and 15.4~GHz, respectively. Subtracting the VLBA flux from the VLA flux
yields
\begin{equation}
F_{\nu}^{LS}(\nu =8.4 \,\rm{GHz}) = 71 \, mJy \; , \;
F_{\nu}^{LS}(\nu =15.4 \,\rm{GHz}) = 49\, mJy \;.
\end{equation}
Fitting the data in Equation (2) by a power law with spectral index
of $\alpha=0.62$, these flux densities can be used to extrapolate an
estimate of the desired result,
\begin{equation}
F_{\nu}^{LS}(\nu =14.6 \,\rm{GHz}) = 51 \, mJy \; , \;
F_{\nu}^{LS}(\nu =17.6 \,\rm{GHz}) = 45\, mJy \;.
\end{equation}
\par The core flux, $F_{\nu}^{\rm{core}}$, can be estimated from the AMI measured flux
density, $F_{\nu}$, by
\begin{equation}
F_{\nu}^{\rm{core}} \approx F_{\nu} - F_{\nu}^{LS} - F_{\nu}^{SS}
\;.
\end{equation}
We can check Equation (4) during the first two epochs of VLBA
observations at 17.6~GHz. Recall that $F_{\nu}^{SS}(\nu
=17.6\rm{GHz}) =$ 74~mJy and 80~mJy from above in epochs 1 and 2.
The extrapolated core flux is $F_{\nu}^{\rm{core}}$ = 90~mJy in both
epochs. Thus, using Equation (3), Equation (4) predicts that
$F_{\nu}(\nu =17.6 \,\rm{GHz}) \approx 45 + 90 + 74$ mJy = 209~mJy
in epoch 1 and $F_{\nu}(\nu =17.6 \,\rm{GHz}) \approx 45 + 90 +
80$~mJy = 215~mJy in epoch 2 which is consistent with the levels
measured (quasi-simultaneously) during plateau 1 in Figure 6. With
the confidence provided by this consistency check, we estimate
$F_{\nu}^{\rm{core}}$ from the AMI light curve in the bottom frame
of Figure 10. Note that the estimated $F_{\nu}(\nu =17.6
\,\rm{GHz})$ increases from 14~mJy to 30~mJy at the time of the HST
observation and the spectral index is extremely inverted and a short
extrapolation to 22.2~GHz implies that $F_{\nu}(\nu =22.2
\,\rm{GHz})\approx 50$~mJy. The bottom frame of Figure 10 indicates
that the nuclear spectrum hardens drastically a few days to 2 weeks
before the G140L observation. A hardening of the spectrum is
expected before the rise of the low frequency flux in the early
stages of an optically thick flare. The implication is that the flux
was increasing at high frequency during this time period. As the
ejected plasma propagates to larger dimension, the SSA opacity
decreases. Around the time of the G140L observations, the spectrum
begins to soften and the SSA opacity is sufficiently low at 17.6~GHz
that the flux density starts to increase rapidly: increasing by a
factor of 4 in less than three weeks.

\par The conclusion is that the elevated flux on 2014 December 15
(MJD 57005.5) is a direct consequence of the flare that peaked 6
months later. Because of the gap in observations between 2014 December 8
and 2014 December 15, the light curve could have begun a
drastic increase at any time in between. Furthermore, for an
optically thick flare at 17.6~GHz, the 43~GHz flux density increase
leads the increase of the 17.6~GHz flux density. Thus, the 2014 December
13 HST COS G140L observation was most likely a few days to two
weeks after the flare initiated.

\begin{figure}
\begin{center}
\includegraphics[width= 0.55\textwidth ]{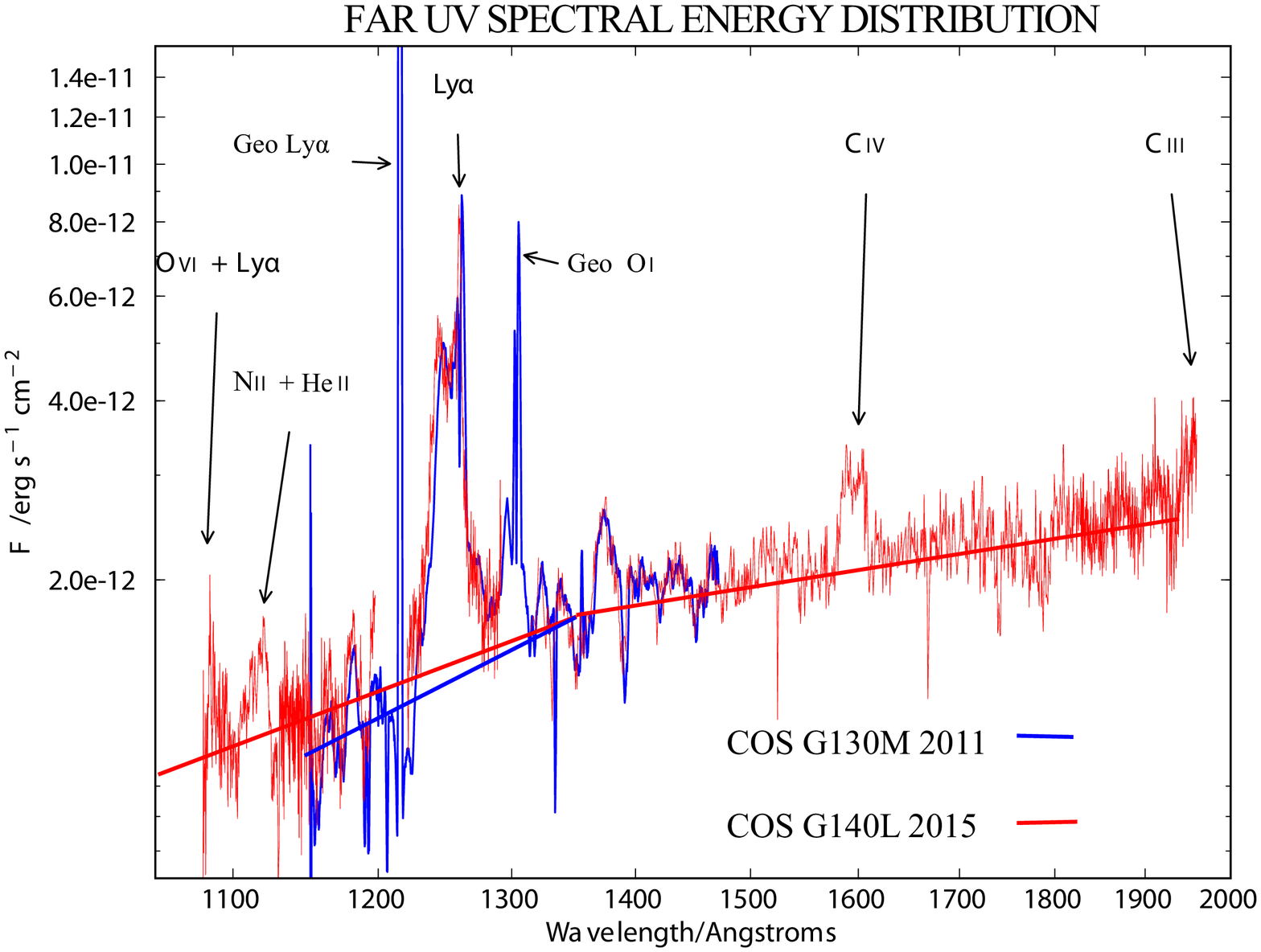}
\includegraphics[width= 0.55\textwidth ]{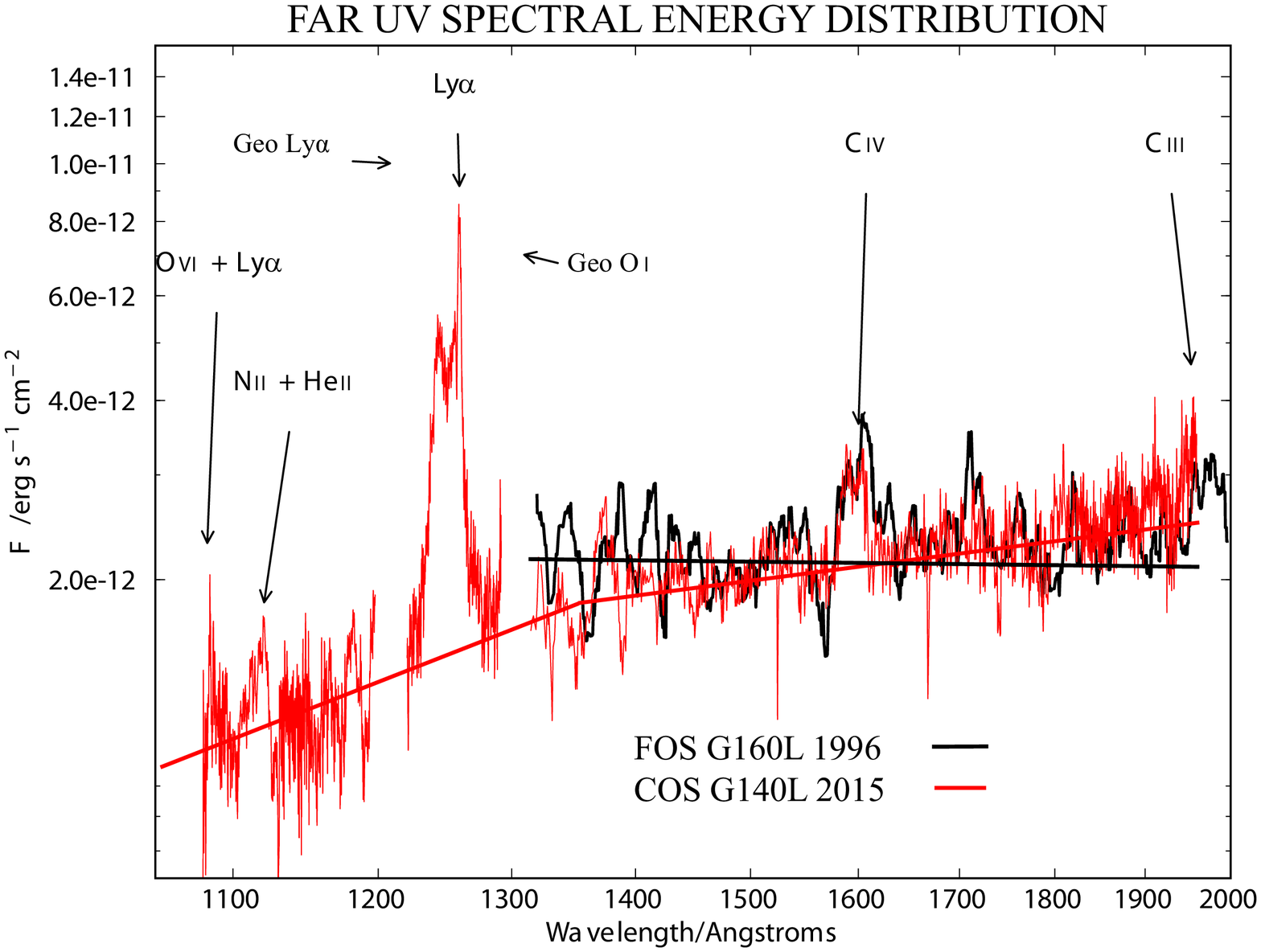}
\includegraphics[width= 0.55\textwidth ]{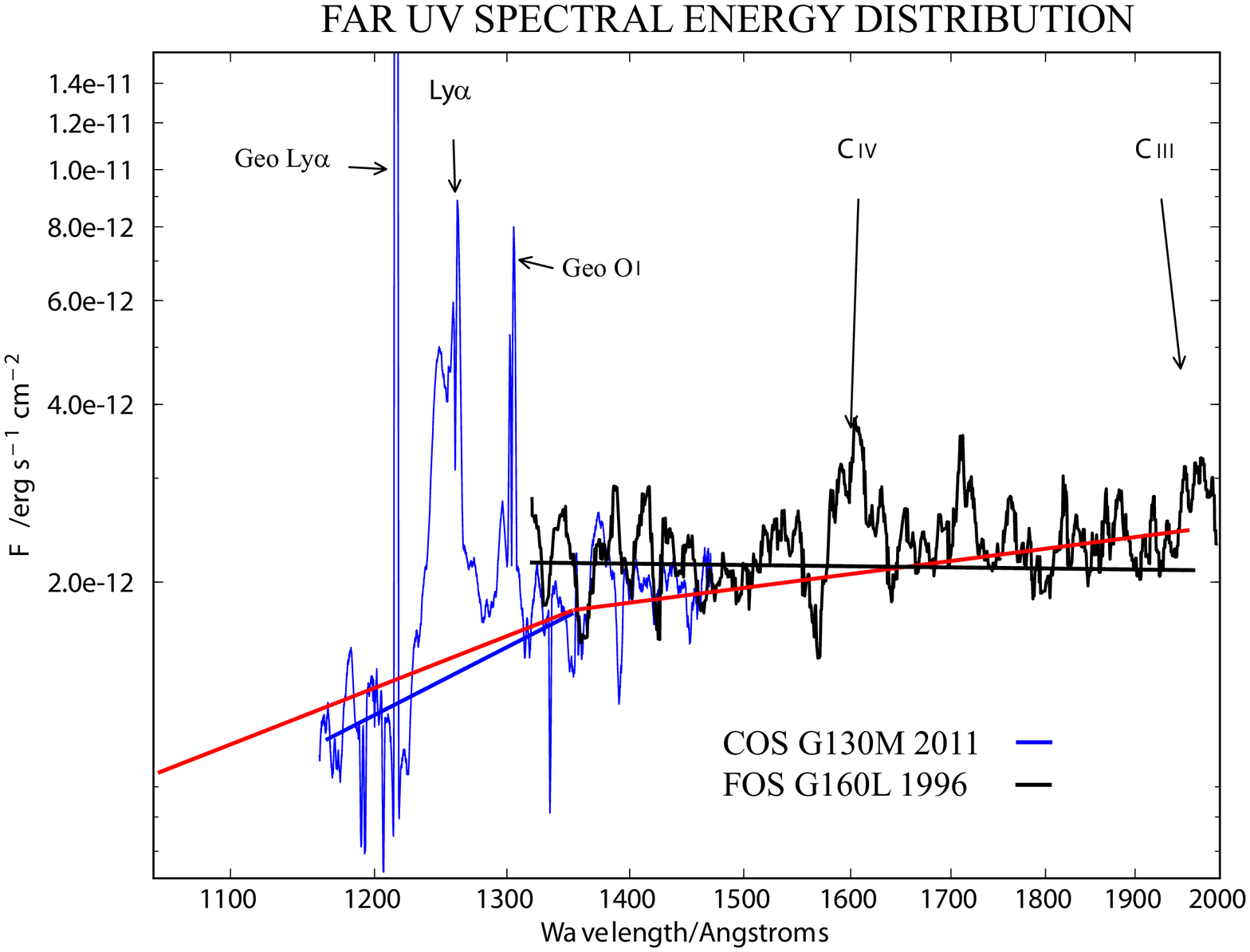}
\caption{The top panel compares the SEDs of the two high radio
states, the 2011 flare and the 2015 flare. The middle panel compares
the SED during the 2015 flare and during the radio low state in
1996. There is CIV BAL absorption and a harder spectrum in 1996. The
last frame compares the SED in 1996 and during the radio flare in
2013. The SEDs diverge shortward of $1400\AA$.}
\end{center}
\end{figure}
\subsection{Previous HST Observations} Another observation that occurred during an active radio
state was the HST COS G130M observation on 2011 October 15 that was
indicated in the top frame of Figure 1. Figure 11 is a magnified
view of the radio flare so that one can see the radio state at the
time of the HST observation. Note that the observation was during a
very high 22~GHz flux state. The flare was experiencing a saddle
point that was either near the peak of the flare or on the declining
side. It appears that both the G140L and the G130M observations
occurred during very active jet states, at the start of a strong
flare and during the last local maximum of another strong flare,
respectively. Conversely, as discussed in terms of Figure~1, the
1996 November 21 HST FOS spectrum was taken near (within two weeks of) an
historic low in the 22~GHz radio flux.
\par There is a
great disparity between the quality of the FOS and COS data-sets.
The highest quality data are the G130M data from 2011. It is averaged
over 10 observations that exceed a total of 12000 s. These data were
previously discussed in \citet{vei13,vei16} and the averaged data
was generously provided by S. Veilleux and M. Melendez. However,
these data had a rather narrow wavelength coverage. The G140L
spectrum from 2014 was downloaded from the Milkulski Archive for the
Space Telescope (MAST) and is of comparable quality with over 15,000
seconds of observation, but a much wider wavelength range and much
lower spectral resolution. These data have been considered extensively
in \citet{vei16}. The FOS G160L data was a short 770s exposure with
low resolution. The following discussion is highly dependent on this
brief observation. The important aspect for this discussion is
knowing if the absolute flux calibration was accurate. In order to
remove ambiguity, we excised the spectrum below $1315 \AA$, in order
to avoid Geo OI coronal emission and Geo Ly$\alpha$ emission that is
impossible to reliably extricate with the low resolution combined
with the noisy edge of the spectral coverage. With these issues
removed, we can assess the flux calibration. The observation was
Post-COSTAR and the acquisition method was ACQ/PEAK that creates a
centering uncertainty  of 0.05" \citep{eva04}. This equates to a 1\%
flux uncertainty in the 1.0" circular aperture (FOS Instrument
Science Reports CAL/FOS-140) that was used. The data acquisition
files had no flags or warnings, so every indication is that the flux
uncertainty should be $<10\%$ \citep{eva04}. The absolute flux
density of the COS observations is likely better. The 10 different
observations with G130M agree within $<5\%$ in the overlapping
ranges of wavelength, similarly for the two observations with G140L.

\begin{table*}
 \centering
\caption{Far UV Spectral Properties Versus High Frequency Radio
Properties} {\tiny
\begin{tabular}{ccccccccccc}
 \hline
1          & 2             &     3          & 4              & 5                  & 6                                           &  7                                          & 8                    & 9                        &  10     \\
Date       & 17.6~GHz     &  22~GHz         & Date           & Grating/          & $F(1365 \pm 50 \AA)$      &  $F(1850 \pm 50 \AA)$    & $\alpha_{1315-1050}$\tablenotemark{a} & $\alpha_{1900-1315}$\tablenotemark{b}   & BALNIcity  \\
           & Flux         &  Flux           &            &          &       &      &  &    &   \\
Radio      &   Density     & Density         &  HST            & Spectro-     & $10^{-15}$                                   &   $10^{-15}$                                &                     &                  & CIV \\
Obs. &   (mJy)     &   (mJy)                 & Obs.            &   graph               & ergs~$\rm{s}^{-1}$~$\rm{cm}^{-2}$                         &   ergs~$\rm{s}^{-1}$~$\rm{cm}^{-2}$                      &                      &                  &  (km/s)      \\

\hline \hline
1996 Dec 05  &  $90 \pm 10$ \tablenotemark{c} & $62\pm9$ & 11/21/1996 & FOS/G160L  & $168 \pm 17$ & $128 \pm 13$ & ... &   $\approx -1$ & $\approx 600$\tablenotemark{d}  \\
2011 Sep 30 &  ...         & $200\pm10$ & 2011 Oct 15 & COS/G130M  & $139 \pm 7$ & ... & 1.7 &   ... & ...  \\
2011 Oct 25  &  ...        & $180\pm5$ & 2011 Oct 15 & COS/G130M  & $139 \pm 7$ & ... & 1.7 &   ... &   ... \\
2014 Dec 15  &  $142 \pm 5$ & ...     & 2014 Dec 13 & COS/G140L  & $140 \pm 7$ & $135 \pm 7$  & 1.5 &   -0.1 & 0\tablenotemark{e}  \\

\hline \hline

\end{tabular}}
\tablenotetext{a}{Spectral index of continuum power law,
$F_{\lambda} \sim \lambda^{\alpha}$. Computed from $1050 \AA$ to
$1300 \AA$.} \tablenotetext{a}{Spectral index of continuum power
law, $F_{\lambda} \sim \lambda^{\alpha}$. Computed from $1315 \AA$
to $1900 \AA$.} \tablenotetext{c}{estimated from the VLA data in
\citet{ulv99}}\tablenotetext{d}{\citet{gal02}}\tablenotetext{e}{\citet{vei16}}

\end{table*}
\subsection{Comparing the Three Epochs of HST Observations} The far UV spectra are compared in Figure 12 by plotting
each pair of spectral energy distributions (SEDs) in different
panels. Plotting all three at once is too cluttered for comparative
purposes. There are some amazing similarities and striking contrasts
that are catalogued in Table 2. The top frame in Figure 12 compares
the two SEDs that were sampled when the radio jet was in a state of
high activity. There is no observable difference in the region of
overlap between $1150\AA$ and $1470\AA$. The Geo OI coronal emission
and Geo Ly$\alpha$ emission were left in for reference in the higher
spectral resolution G130M SED. These regions were excised from the
G140L SED for clarity. The spectral index of $\alpha_{\lambda}=1.7$
in column (8) of Table 2 for the continuum flux density,
$F_{\lambda} \sim \lambda^{\alpha}$, between $1315 \AA$ and $1050
\AA$ agrees with the value found previously for the G130M spectrum
\citep{vei13}. Formally, the spectral coverage ends at $1150 \AA$,
so the last $100\AA$ is an extrapolation. The G140L spectrum shows a
spectral break near $1315 \AA$.
\par The middle frame compares the high activity jet state at the
beginning of the 2015 flare and the 1996 low radio luminosity state.
The SED is very similar from $1400\AA$ to $1900\AA$, even after 18
years. However shortward of $1400\AA$ there is a clear visual excess
in the 1996 SED. We would like to quantify what we see visually. One
method was to compute the integrated flux in two bands, $F(1365 \pm
50 \AA)$ in column (6) of Table 2 and in the range $F(1850 \pm 50
\AA)$ in column (7) of Table 2. Even though the flux in the band
$F(1850 \pm 50 \AA)$ is slightly larger in 2014, it is significantly
smaller in the band $F(1365 \pm 50 \AA)$ in 2014. Visually it is
very clear in the figure that the G140L SED angles downward toward
shorter wavelength between $1315 \AA$ and $1900 \AA$. However, the
data are inconsistent with such a downward slant in 1996. We quantify
this in Table 2 in Column (9) with the spectral index
$\alpha_{\lambda} \approx -1$ in 1996 (a flat SED) compared to
$\alpha_{\lambda} \approx -0.1$ in 2014. The other significant
distinction is the depth of the CIV BAL trough. The BALnicity index,
BI, was defined in \cite{wey91} as
\begin{equation}
BI = \int_{v=-25000}^{v=-3000}  [1 -F(v)/0.9]C\, dv\;,
\end{equation}
where $F(v)$ is the flux density normalized to the continuum level
as a function of $v$, the velocity from the QSO rest frame line
emission frequency in km/s. The step function, $C(v) \neq 0$ if and
only if there is more than 2000 km/s of continuous absorption beyond
-3000 km/s. This measure of broad absorption has been borne out over
the years as very robust. More lax measures of absorption (such as
associated absorption and mini-BALs) have turned out to represent
different classes of objects, that have less X-ray absorption than
BALQSOs and larger radio luminosity than bona-fide BALQSOs
\citep{sha08,pun06,kni08}. Column (10) of Table 2 lists the
estimated values of BI. During the low radio state in 1996, $BI
\approx 600$ km/s \citep{gal02}. At the beginning of the 2015 flare
$BI =0$ km/s \citep{vei16}. In the middle panel of Figure 12, it is
evident that there is more high velocity absorption in 1996. \par
The bottom panel of Figure 12 compares the SED during the 2011 flare
and the quiescent state in 1996. Even though the overlap in spectra
is only from $1315\AA$ to $1470\AA$, it is clear that the SEDs agree
in the range of $1400\AA$ -- $1470\AA$, but they clearly diverge
shortward of this. The long wavelength end of the G140L continuum
fit is left on the plot for an additional reference level.
\par In summary, based on limited data and the inability to secure
an HST spectra during the peak of the 2015 as a target of
opportunity, we find some trends that would need to be verified with
more spectra.
\begin{figure}
\begin{center}
\includegraphics[width= 0.8\textwidth ]{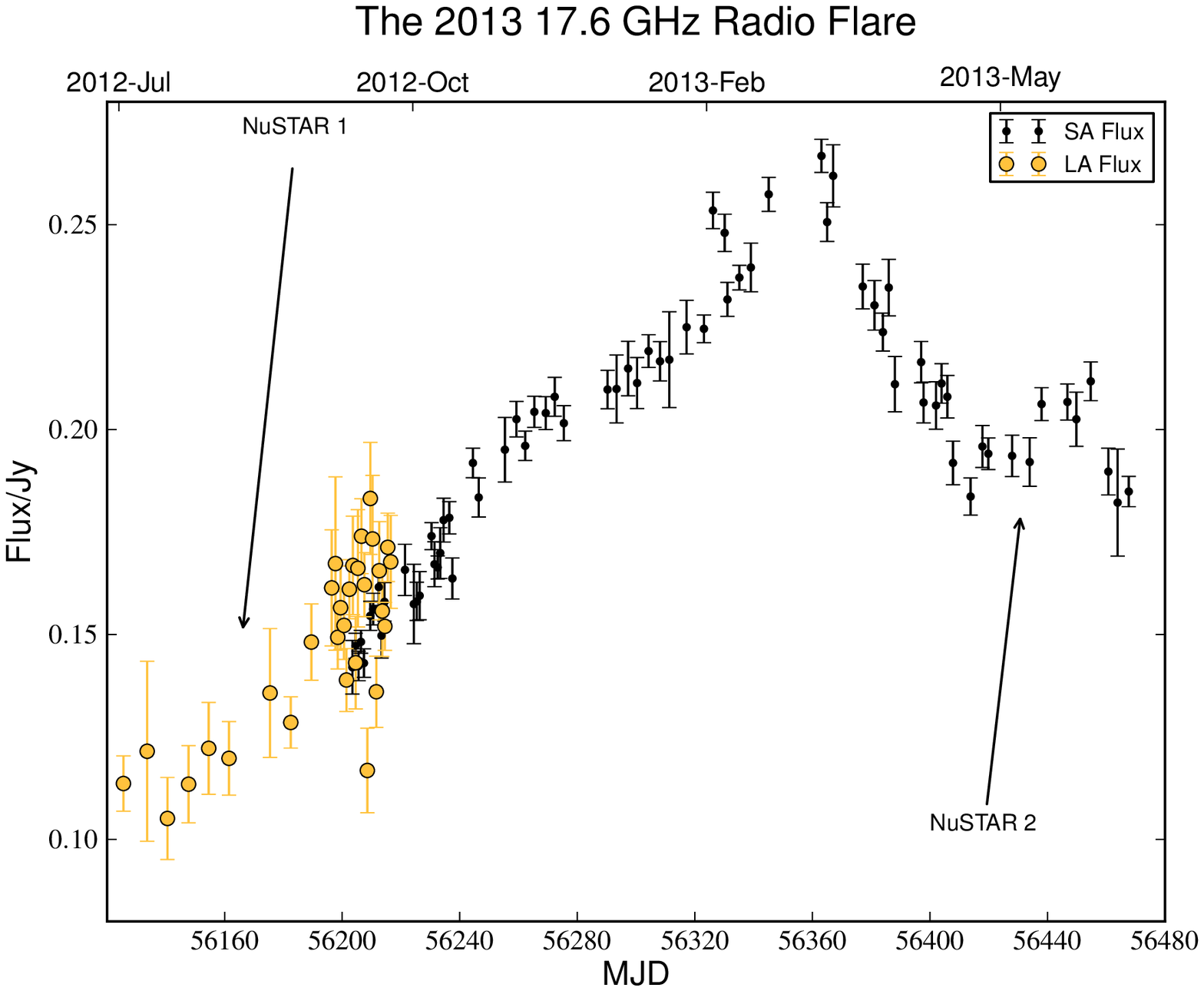}

\caption{The light curve of the 2013 flare shows that the NuSTAR 1
observation occurred during a low radio state and the NuSTAR 2
observation occurred during a high radio state, As discussed in
\citet{rey13}, the AMI monitoring began in the large array. However,
as can be seen from the onset of some large scatter, calibration
issues developed and we shifted to the small array.}
\end{center}
\end{figure}
\begin{enumerate}
\item The far UV emission longward of $1400\AA$ is incredibly
constant over an 18 year period. This finding is unusual for a
quasar. Studies of far UV variability of quasars indicate measurable
variability is the norm for these time frames although no
variability is occasionally found \citep{mac12,wel11,wil05}.
\item There is a far UV flux deficit shortward of $1400\AA$ during
periods of increased relativistic jet activity, relative to the far
UV flux in periods of low radio activity.
\item The high ionization BAL absorption is suppressed during
periods of increased relativistic jet activity.
\item There is a sharp spectral break near $1400\AA$ during periods
of high jet activity. It is unknown what happens during periods of
low radio emission. The peak of the far UV SED is longward of
$1400\AA$. The steep spectral index shortward of $1400\AA$,
$\alpha_{\lambda}=1.5 - 1.7$, is extremely steep and unusual for a
quasar \citep{tel02}.
\end{enumerate}

\section{X-ray Observations}

\begin{table}
\caption{Log of the X-ray observations used in this work.}
\label{Xrayobs}
\begin{center}
\begin{tabular}{l c c c c}
\hline\hline
Mission & Name &Date & Detector & Net exposure \\
\hline \hline
XMM-Newton  & XMM 1 &  2001 Jun 07   & EPIC-pn & 17~ks   \\
\hline
NuSTAR  &  NuSTAR 1 & 2012 Aug 27  & FPM~(A-B) & 41~ks   \\
\hline
NuSTAR  & NuSTAR 2 &  2013 May 09   & FPM~(A-B) & 29~ks   \\
\hline
&& &    \\
\hline
NuSTAR  & NuSTAR 3 &  2015 Apr 02   & FPM~(A-B) & 32~ks   \\
\hline
NuSTAR  & NuSTAR 4 &  2015 Apr 19   & FPM~(A-B) & 28~ks   \\
\hline
XMM-Newton  & XMM 2 &  2015 Apr 25   & EPIC-pn & 19~ks   \\
\hline
NuSTAR  & NuSTAR 5 &  2015 May 28   & FPM~(A-B) & 30~ks   \\
\hline
XMM-Newton  & XMM 3 &  2015 May 28   & EPIC-pn & 21~ks   \\
\hline  \hline
\end{tabular}
\end{center}
\end{table}

Mrk\,231 was observed three times with NuSTAR and twice with
XMM-Newton during the 2015 campaign. As reported in
Table~\ref{Xrayobs}, NuSTAR observed the source on April 2 ($\sim
32$~ks of net exposure), on April 19 ($\sim 28$~ks), and on May 28
($\sim 30$~ks), while the two XMM-Newton observations were carried
out on April 25 ($\sim 19$~ks), and on May 28 ($\sim 21$~ks).
Table~\ref{Xrayobs} also reports one previous XMM-Newton observation
(performed on 2001 June 7) and two archival NuSTAR observations
performed on 2012 August 26 and 2013 May 9 that we use for
comparison purposes.

\subsection{The 2015 {\it NuSTAR} Observations}

In order to assess the relevance of the five pointed observations of
Mrk\,231 with NuSTAR in the context of jet activity, we consult the
light curve in Figures 1 and 2. More resolution is needed in Figure
1 in order to clarify the radio state for the NuSTAR 1 observation.
The light curve in Figure 13 of the 2013 flare indicates that the
NuSTAR 1 observation occurred during a low radio state and the
NuSTAR 2 observation occurred during a high radio state (as did the
NuSTAR 3,4, and 5 observations in 2015). Thus, NuSTAR 1 is in a low
radio state and NuSTAR 2--5 are all in a high radio state.

\begin{figure}
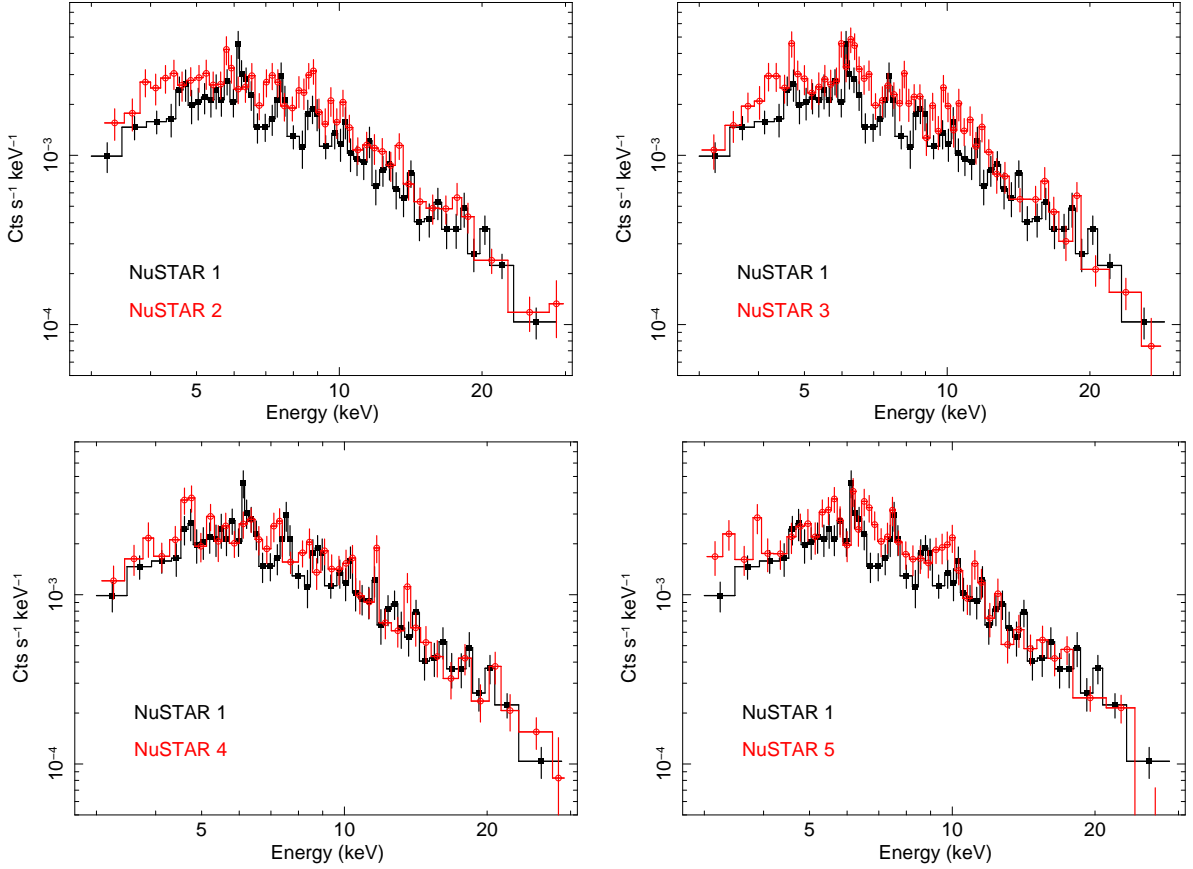

  \begin{center}
    \mbox{\includegraphics[width= 0.34\textwidth,angle=-90]{NuSTAR1-2comp_lines.ps}
      \includegraphics[width= 0.34\textwidth,angle=-90]{NuSTAR1-3comp_lines.ps}}
    {\vspace{0.2cm}}
    \mbox{\includegraphics[width= 0.34\textwidth,angle=-90]{NuSTAR1-4comp_lines.ps}
        \includegraphics[width= 0.34\textwidth,angle=-90]{NuSTAR1-5comp_lines.ps}}
    \caption{The low radio flux state NuSTAR 1 X-ray spectrum (black
      filled squares in all panels) is compared with all other high
      radio flux NuSTAR spectra (red empty circles in all panels) in
      the most sensitive NuSTAR band (3--30~keV). Each spectrum is the
      result of co-adding the FPM A and B NuSTAR detectors, as this
      improves visual clarity. The co-added spectra are produced by
      combining source and background spectra, as well as responses
      using the {\it addspec} FTOOL available within the HEASOFT
      package. Solid lines are plotted only to guide the eye, and they
      step through all data points (no spectral fit is included
      here).}
    \label{NuSTARcomp}
  \end{center}
\end{figure}

The NuSTAR 1 low radio state X-ray spectrum is compared with all
other high radio flux NuSTAR observations in
Figure~\ref{NuSTARcomp}. We note that the NuSTAR 1 observation
presents a clear deficit of X-ray photons below 12~keV with respect
to the NuSTAR 2, 3 and 5 observations. This is especially noticeable
bluewards of the Fe emission line (around 6~keV in the observed
frame) where there are hints for absorption structures in NuSTAR 1
between 6.5~keV and 10~keV. The NuSTASR 1 and NuSTAR 4 spectra seem
to be more similar, although some subtle difference may be present
in two narrow bands areound 7~keV and 8.5~keV. On the other hand,
the high-energy spectra above $\sim$~12~keV look the same at all
epochs. These qualitative claims are confirmed by comparing the
(model-independent) background-subtracted count rates in the soft
and hard bands for all observations. The net count rates are
reported in Table~\ref{CtsTable}, and they confirm that the NuSTAR 1
observation is characterized by a deficit of soft X-ray photons with
respect to the NuSTAR 2, 3, and 5 pointings. NuSTAR 4 appears to
have a soft X-ray deficit similar to NuSTAR 1, while high-energy
count rates above 12~keV are consistent with each other at all
epochs, with perhaps a hint of decreased flux in the last NuSTAR
observation. This suggests that any difference in the X-ray spectrum
between epochs is likely due to intervening absorption rather than
being intrinsic. The shape of the NUSTAR 1 spectrum between 6.5~keV
and 10~keV seems indeed to suggests that some absorption feature is
present in that band. This structure may be present in other
observations as well, but with reduced strength.

\begin{table}
\caption{The net count rates for all NuSTAR observations in the
  3--12~keV and 12--30~keV bands. Count rates are derived from the
  co-added A+B detectors data for all observations. Units are $10^{-2}$~cts~s$^{-1}$.}
\label{CtsTable}
\begin{center}
\begin{tabular}{l c c }
\hline\hline
Observation & 3--12~keV count rate & 12--30~keV count rate\\
\hline \hline
NuSTAR 1 & $1.44\pm 0.05$ & $0.56\pm 0.03$\\
NuSTAR 2 & $1.89\pm 0.06$ & $0.61\pm 0.04$\\
NuSTAR 3 & $1.97\pm 0.06$ & $0.57\pm 0.03$\\
NuSTAR 4 & $1.62\pm 0.06$ & $0.54\pm 0.04$\\
NuSTAR 5 & $1.79\pm 0.06$ & $0.52\pm 0.03$\\
\hline  \hline
\end{tabular}
\end{center}
\end{table}

We then proceed by fitting all available NuSTAR data in the highest
signal-to-noise band, namely the 3 -- 30 keV one. Mrk\,231 is
unresolved by the NuSTAR beam, thus star formation and high mass
X-ray binary emission are responsible for a non-negligible fraction
of the total X-ray emission detected by NuSTAR \citep{ten14}. We do
not want to speculate on complicated models with many uncertainties
and assumptions and we use here some very simple spectral models.
The baseline model is that of a simple power law continuum absorbed
by a Galactic column density fixed at $9.64\times
10^{19}$~cm$^{-2}$, \citet{kal05}) plus an extra absorbing column of
neutral gas at the redshift of the source, and a Gaussian emission
line with $\sigma$ fixed to $10$~eV (unresolved at the resolution of
the NuSTAR A and B detectors) to reproduce the Fe K emission line
(the line energy is restricted between 6.4~keV and 6.97~keV in the
rest-frame). All our spectral analysis is first carried out
considering the FPM A and B NuSTAR detectors separately, including a
cross-normalization constant between the two. We then produce
co-added spectra (A+B) for all observations, and we re-fit the
co-added spectra with the same spectral model. As we do not find
any statistically significant difference in any of the spectral
parameters we derive, we proceed by reporting results for the
co-added A+B spectra.

\begin{figure}
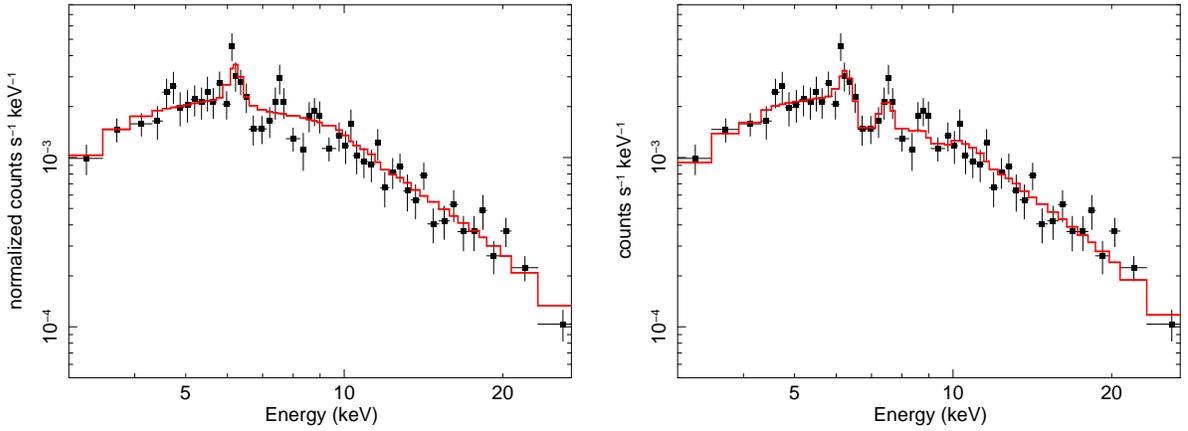

  \begin{center}
\includegraphics[width= 0.34\textwidth,angle=-90]{N1_AandB_simple.ps}
\includegraphics[width= 0.34\textwidth,angle=-90]{N1_AandB_phase.ps}
    \caption{The NuSTAR 1 X-ray spectrum and best-fitting baseline
      model are shown in the left panel. The right panel presents the
      same data together with the best-fitting model including an
      ionized X-ray wind that accounts well for the deficit of
      photons in the 6.5--10~keV band (observed frame).}
    \label{NuSTAR1}
  \end{center}
\end{figure}

All of the NuSTAR observations except for the low radio state NuSTAR
1 observation are well fitted by the simple baseline model with
$\chi^2$ of the order of unity. NuSTAR 1 represents the exception,
and the baseline model provides a poor fit to the data with
$\chi^2=82$ for 56 degrees of freedom (dof). The second worse fit is
obtained for NuSTAR 4 (as expected given that this is the
observation that mostly resembles NuSTAR 1, see
Figure~\ref{NuSTARcomp}) where we have $\chi^2 = 58$ for 40 dof. Our
results are reported in Table~\ref{NuSTARbaseline}, and the peculiar
NuSTAR 1 spectrum and best-fitting baseline model are shown in the
left panel of Figure~\ref{NuSTAR1}.

\begin{table*}
\caption{Baseline model fits to all NuSTAR observations. Column
  densities are expressed in units of $10^{22}$~cm$^{-2}$, the Fe line
  energy is in keV, and its intensity is in units of
  $10^{-6}$~ph~s$^{-1}$~cm$^{-2}$. Note that the Fe line is not
  formally detected during the NuSTAR 2 and 4 observations (we fix its
  energy to 6.6~keV in order to compute an upper limit on its
  intensity).}
\label{NuSTARbaseline}
\begin{center}
\small{\begin{tabular}{l c c c c c} \hline\hline
Observation & $\Gamma$ & $N_{\rm H}$ & $E_{\rm{Fe}}$ & $I_{\rm{Fe}}$ & $\chi^2$/dof \\
NuSTAR 1 & $1.20\pm 0.20$ & $5 \pm 4$ & $6.5\pm 0.1$ & $3.5\pm 2.0$ & 82/56  \\
NuSTAR 2 & $1.50\pm 0.20$ & $5 \pm 4$ & $6.6^f$ & $\leq 2.5$ & 47/45  \\
NuSTAR 3 & $1.60\pm 0.20$ & $10 \pm 5$ & $6.55\pm 0.15$ & $4.4\pm 2.3$ & 54/52  \\
NuSTAR 4 & $1.40\pm 0.20$ & $5 \pm 4$ & $6.6^f$ & $\leq 3.5$ & 58/40  \\
NuSTAR 5 & $1.50\pm 0.20$ & $5 \pm 4$ & $6.7\pm 0.2$ & $2.8\pm 1.8$ & 54/45  \\
\hline  \hline
\end{tabular}}
\end{center}
\end{table*}

It is interesting to note that the only data set that is not well
described by the baseline model (the NuSTAR 1 low radio state
spectrum in 2012) was used, together with a simultaneous and higher
quality Chandra X-ray spectrum (and to the NuSTAR 2 data) to argue
that an ultra-fast wind is present in Mrk\,231. The evidence for such
wind was detected in X-ray absorption from 6.5 keV to 10 keV
\citep{fer15}. This appears to be a likely explanation for the
deficit of X-ray photons in the 6.5-10~keV band during the NuSTAR 1
observation (see the left panel of Figure~\ref{NuSTAR1}) although
our analysis seems to exclude that the NuSTAR 2 observation (high
radio state) can be interpreted in a similar way. This is unlikely
to represent a problem for the claimed wind detection, as the
overall statistical result is dominated by the Chandra spectrum,
simultaneous with NuSTAR 1. As mentioned, from a radio point of
view, the NuSTAR 1 and Chandra observations were performed during a
low radio state, while NuSTAR 2 was already in a high radio state.
As the NuSTAR 2 spectrum is well described by the baseline model and
does not exhibit signs for a deficit of photons in the interesting
band, it is tempting to associate the presence of the wind with low
radio states.

In order to see whether the peculiarity of the NuSTAR 1 spectrum is
indeed well described by an X-ray wind, we add a further ionized
absorber to our spectral model for the NuSTAR 1 observation. The
additional ionized absorber (modelled via the photoionization code
PHASE, see \citet{kro03}) depends on the absorber column density,
ionization, and outflow velocity as well as on the gas turbulent
velocity. The latter parameter is however fixed to an intermediate
value of $500$~km~s$^{-1}$ as the poor spectral resolution of the
NuSTAR data are not sufficient to constrain it.

The ionized absorber provides a statistical improvement of the fit
for the NuSTAR 1 data with $\chi^2 = 65$ for 53 degrees of freedom
(dof), to be compared with $\chi^2 = 82$ for 56 dof of the baseline
model, i.e. the ionized absorber is detected at the 99.4 per cent
confidence level according to a simple F-test. The addition of the
ionized absorber produces a slightly steeper photon index for the
NuSTAR 1 spectrum, which is more in line with that from all
others ($\Gamma=1.4\pm 0.2$ versus $\Gamma=1.2\pm 0.2$ with the
baseline model). All other parameters are the same, within the
errors, as those obtained with the baseline model. From our
best-fitting model, we infer that the ionized absorber has a column
density of $(5\pm 3)\times 10^{23}$~cm$^{-2}$ with ionization
parameter $\log U = 2.3\pm 0.3$, and that the gas is outflowing from
the system at high velocity, namely v$_{\rm out} = 17000\pm
4500$~km~s$^{-1}$. The NuSTAR 1 spectrum and best-fitting model are
shown in the right panel of Figure~\ref{NuSTAR1}. We note that the
wind parameters that we derive from the NuSTAR 1 data are all
consistent within the errors with those reported by other authors
who made use of a simultaneous Chandra observation as well
\citep{fer15}. This is reassuring in terms of the reality of the
wind detection in the NuSTAR 1 observation.

We then perform similar fits including a putative ionized outflow to
all other NuSTAR observations, but we do not find any statistical
evidence for the presence of such component in any of the high radio
state NuSTAR observations (NuSTAR 2--5). This is true also for the
NuSTAR 4 observation despite the fact that its spectral shape
appears to be the most similar to NuSTAR 1 (see
Figure~\ref{NuSTARcomp}), and we only obtain an improvement of
$\Delta\chi^2 = 5$ for 3 dof for the NuSTAR 4 observation when a
wind model is included. Obviously, this does not mean that the wind
is absent during the NuSTAR 2--5 observations, but only that its
presence is not required by the data (using our specific wind
model). The only significant detection is thus obtained during the
only low radio flux state observation NuSTAR 1 where it explains
well the observed deficit of soft X-ray photons below 12~keV. If
the flux deficit during the radio weak state is indeed due to an
X-ray absorbing wind then this is perhaps suggestive of a phenomenon
known to occur in Galactic black hole accreting systems, where jet
production is seen to suppress winds of X-ray absorbing gas
\citep{nei09}. It would also provide an interesting analogy to the
suppression of BAL winds in radio loud states noted for Mrk\,231 in
the last section.

\subsection{The 2015 XMM-Newton observations}

Mrk\,231 has been observed three times with XMM-Newton, twice in
this observing program and once in 2001. As can be seen from the
light curve in Figure 1, the radio state is unknown during the 2001
XMM-Newton observation \citep{bra04}. So we have two observations
during active jet states and one during an unknown epoch of jet
activity. Figure 16 shows that there is no measurable difference
between the three spectra obtained with XMM-Newton.

Even though the 2015 XMM-Newton spectra are redundant with the 2001
spectrum, we proceed with a simple spectral analysis of these data.
XMM-Newton and NuSTAR observed Mrk\,231 simultaneously on 2015 May
28. We then start our analysis with this data set, performing
simultaneous fits to the low and high energy data from the two
missions, to gain insights on the overall broadband spectrum of
Mrk\,231. A constant is introduced to account for any
cross-calibration issue between the two missions. The XMM-Newton
EPIC-pn data (XMM 3) are considered in the 0.5--12~keV band, while
the NuSTAR 5 data (A+B detectors) are used in the 3--30~keV band as
above. The soft X-ray emission below $\sim 2$~keV is dominated by
extended emission (star-formation, photo-ionized gas, etc.) and it
can be successfully modeled by means of a two-temperature thermal
plasma model plus a power law component representing scattered
emission. These two components are thought to originate far away
from the central engine, and are thus only absorbed by the Galactic
column.  As for the $\geq 2$~keV spectrum, several models are
possible (see e.g. \citet{pic13}) and we start here from the
simplest possible one, namely an extra absorption component at the
redshift of Mrk\,231 acting on a power law continuum plus Fe line
(i.e. the same baseline model that was applied to the NuSTAR data in
the previous subsection).

This first attempt provides a good statistical result with
$\Delta\chi^2 = 170$ for 134 dof. The temperature of the hot plasmas
is $\simeq 0.24$~keV and $\simeq 0.89$~keV, and this component
contributes to the soft 0.5--2~keV band with a total luminosity of
$\simeq 2\times 10^{41}$~ergs~$\rm{s}^{-1}$. The soft (scattered)
power law has a loosely constrained photon index $\Gamma = 1.5\pm
0.4$ and contributes to the 0.5--2~keV luminosity with an additional
$\simeq 2.5\times 10^{41}$~ergs~$\rm{s}^{-1}$. As for the
high-energy emission, the X-ray nuclear continuum is well described
by a power law with $\Gamma = 1.45\pm 0.15$ with an intrinsic
2--10~keV (10--30~keV) luminosity of $\simeq 3.0 \times
10^{42}$~ergs~$\rm{s}^{-1}$ ($\simeq 4.2 \times
10^{42}$~ergs~$\rm{s}^{-1}$). If the spectral model is extrapolated
down to 0.5~keV, the total 0.5--30~keV intrinsic luminosity of the
AGN turns out to be $\simeq 8.4 \times 10^{42}$~ergs~$\rm{s}^{-1}$.
However, as the intrinsic continuum is in fact detected only above a
few keV due to intervening absorption, we caution that this value
should be treated with caution as we ignore the true AGN spectrum in
the 0.5--2~keV band (where standard AGN often exhibit a soft X-ray
excess). The neutral absorber has a column density of $6\pm 3\times
10^{22}$~cm$^{-2}$, consistent with that derived from the NuSTAR 5
data alone ($5\pm 4\times 10^{22}$~cm$^{-2}$). We detect an Fe
emission line at $6.6\pm 0.1$~keV with EW$\sim 180$~eV. An emission
feature appears to be present around 5.7~keV in the XMM 3 and, with
less contrast, NuSTAR 5 spectra. The addition of a further emission
line with $\sigma$ fixed at 10~eV results in a statistical
significant improvement of $\Delta\chi^2 = -15$ for 2 dof for a line
energy of $5.7\pm 0.1$~keV and an equivalent width of $\sim 190$~eV.
No significant absorption feature that may indicate the presence of
an ionized wind is seen in the XMM 3 data, as already noted in the
analysis of the NuSTAR 5 data alone.

\begin{figure}
  \begin{center}
    \label{XMMcomp}
\includegraphics[width= 0.5\textwidth,angle=-90]{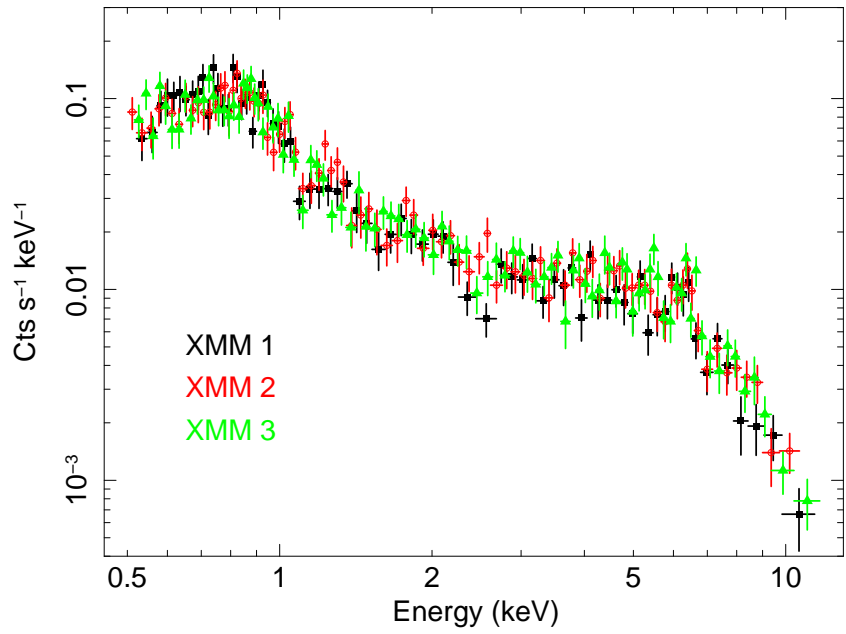}
\caption{The EPIC-pn spectrum from the three epochs of XMM-Newton
  observations show no significant spectral/flux variability.}
\end{center}
\end{figure}

The previous XMM 2 observation that took place on 2015 April 25 can
be described with a very similar model. The only noticeable
difference is that the Fe band appears to be simpler than on May 28.
A simple Fe emission line at $6.7\pm 0.1$~keV with equivalent width
of $\sim 200$~eV describes well all the emission, casting some
doubts on the reality of the $5.7$~keV emission line detected on May
28. The intrinsic continuum photon index is very loosely constrained
by the XMM-Newton data alone ($\Gamma=1.45\pm 0.55$) and it is
absorbed by a column density of $(7\pm 4)\times 10^{22}$~cm$^{-2}$.
The resulting best-fitting statistics is $\Delta\chi^2=79$ for 87
dof, and no features indicate the presence of a further ionized
absorber during the XMM 2 observation.

\subsection{Summary of the X-ray -- Radio Connection}

The X-ray spectrum of Mrk\,231 has been very constant over all
epochs. Because of X-ray absorption of the nuclear emission, the
very soft X-ray emission below a few keV is dominated by extended
components (star-forming regions, scattering, re-emission by
photoionized gas, etc.) and the stability of the X-ray emission is
expected. On the other hand, the remarkably constant X-ray flux
(and spectral shape) above 10~keV indicates that the intrinsic
nuclear emission is not varying on timescales of years.  It is
curious how the far UV and X-ray spectra tend to be very stable,
quite unusual for a broad line active galactic nucleus. The only
epoch where an elevated flux density above 10~keV was seen was with
Suzaku \citep{pic13}, where it was explained as a transient
uncovering of the X-ray nucleus (no radio sampling is available
around the Suzaku epoch).

From a radio point of view, besides the XMM 1 observation that was
performed during an unknown radio state, the NuSTAR 1 pointing is
the only one that is performed during a low radio flux state, while
all others (NuSTAR 2--5 and XMM 2--3) are obtained during high radio
flux states. It is interesting to note that the NuSTAR 1 spectrum is
precisely the only one that cannot be well described by the simple
spectral model used for all other observations. The NuSTAR 1
spectrum presents a deficit of relatively soft X-ray photons
(mostly in the 6.5--10~keV band), while it is indistinguishable from
all others at higher energies, where the intrinsic nuclear emission
dominates. This peculiar spectral shape is best explained by the
addition of a further X-ray absorption component in the form of a
fast ionized wind with outflow velocity $v_{\rm{out}} \sim
17000$~km~s$^{-1}$. Such a wind solution is not statistically required
by the data in any of the high radio state observations. The
presence of an ultra-fast X-ray wind in August 2012 was claimed
previously by others who made additional use of higher quality
Chandra data, simultaneous with NuSTAR 1 \citep{fer15}. The wind
parameters that we derive from the NuSTAR 1 data alone are in
excellent agreement with those previously derived, which indicates
that our detection and modeling are robust.

There being only one low radio state data point, we do not know if the
presence of the wind is related to suppressed jet emission, but it
is suggestive to note that our findings are in line with the
phenomenology of Galactic black hole binaries where winds and jets
seem to be mutually exclusive. If the X-ray wind was indeed present
(or stronger) only during periods of suppressed jet activity, it
would also link well with the suppression of BAL winds during high
radio flux states, as noted in the previous section.

\section{Conclusion}
In this article we used our long term radio monitoring at high
frequency and a target of opportunity monitoring of a strong radio
flare with VLBA, {\it XMM-Newton} and {\it NuSTAR} in order to
probe the radio jet-accretion disk connection in Mrk\,231.

\subsection{Results}In spite of some information loss due to a poor beam shape
in two of our epochs, we were able to determine a few interesting constraints
on the jet-disk connection in this quasar.
\begin{enumerate}
\item As discussed in Section 3, two new ejected components in Mrk\,231 were detected for the
first time (see Figures 4 and 7)
\item In Section 3.2, we were able to constrain the apparent speed of one
of these ejections $ > 3.15$c.
\item In section 3.3, we showed in Figure 8 that the stationary
secondary component (the putative radio lobe) doubled in brightness
in 9 years.
\item In Section 4, we determined that the far UV emission beyond the
peak of the far UV SED was depressed in the two active jet states
compared to the one quiescent jet state monitored with HST (see
Figure 12 and Table 2).
\item In Section 4, we demonstrated that the degree of CIV broad line absorption
in the two active jet states is suppressed compared to the one
quiescent jet state monitored with HST (see Figure 12 and Table 2).
\item In Section 5, we found that the 3 -- 12 keV flux is elevated
during a state of high radio jet activity relative to the one 3 -- 12
keV flux measurement during the only X-ray observation known to
occur in a quiescent radio state (see Figure 14).
\end{enumerate}

\subsection{Physical Interpretation of Results}
\begin{figure}
  \begin{center}
\includegraphics[width= 0.85\textwidth,angle=0]{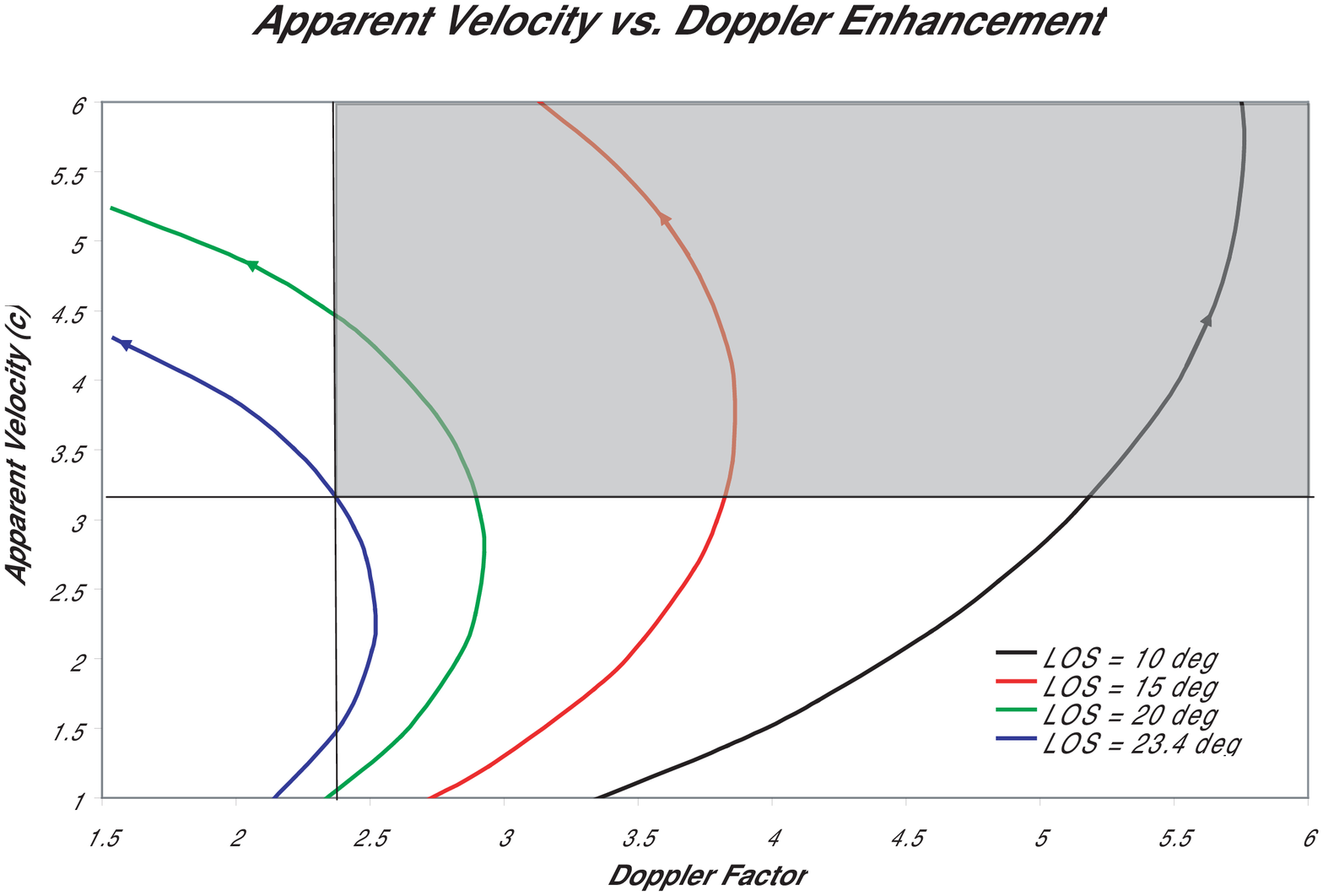}
\caption{The apparent velocity versus Doppler factor for various
lines of sight (LOS). The shaded grey area is consistent with the
lower bounds on the time variability Doppler factor from
\citet{rey09} and $v_{\rm{app}}$ from this paper. The arrows
indicate the direction of increasing $\Gamma$ along each curve.}
\end{center}
\end{figure}
The primary result of this VLBI observation program was the
detection of ejected components and estimating the apparent
velocity, $v_{\rm{app}} > 3.15$c. This allows us to attain new
kinematic insight into the jet. Previously, using the methods of
\citet{gho07}, the time variability brightness temperature is $T_{b}
= (12.4 \pm 3.5)\times 10^{12}$ K \citep{rey09}. In order to avoid
the inverse Compton catastrophe, the analysis in \citet{gho07}
indicates that the line of sight to the jet is less than
$\theta_{max}= (25.6^{\circ})^{+3.2^{\circ}}_{-2.2^{\circ}}$. This
equates to a time variability factor $\delta > (T_{b}/ 10^{12}
K)^{1/3} = 2.3$ for an optically thick, unresolved core
\citep{gho07}. The Doppler factor, $\delta$, is given in terms of
$\Gamma$, the Lorentz factor of the outflow; $\beta$, the three
velocity of the outflow and the angle of propagation to the line of
sight, $\theta$; $\delta=1/[\Gamma(1-\beta\cos{\theta})]$
\citep{lin85}. Figure 17 combines our previous findings with the new
estimate of the component apparent velocity, $v_{\rm{app}}/c = \beta
\sin{\theta}/(1-\beta\cos{\theta})$. The figure plots
$v_{\rm{app}}/c$ as a function of $\delta$. The theoretical plots
are made for various lines of sight (LOS) that are allowed
kinematically. We now have two constraints on $\delta$ and
$v_{\rm{app}}/c$ if we assume a steady jet structure over time. The
intersection of the two lower bounds is shaded in grey. The arrows
indicate the direction of increasing $\Gamma$ on each curve. It is
interesting that the new constraint does not significantly improve
the previous limit on the LOS, $\theta_{max} \approx 23.4^{\circ}$.
The estimate of $v_{\rm{app}}/c$ essentially indicates consistency
with the time variability Doppler factor analysis. Combining the two
conditions in Figure 17 does indicate that $\Gamma > 3.5$ which is a
new constraint on the jet kinematics. This highly relativistic
motion is consistent with a point of jet origin deep within the
gravitational potential well, possibly the innermost regions of the
accretion disk.
\par In principle, polarimetry can provide useful diagnostics of jet
activity. Unfortunately, in \citet{rey09}, it was found that there
was no detectable polarized flux in our VLBA observations.
Presumably, the emission is depolarized by a large scattering column
density during the radiative transfer through the enveloping dense
nuclear environment. Consequently, the 2015 observations were not a
polarization experiment, there was no polarization calibrator in
order to maximize the amount of time on the source. Even so, we
looked for evidence of polarization and could not find any.

Since, the relativistic jet has a plausible origin in the inner
accretion disk there is a straightforward interpretation of point 4,
above. A deficit of far UV emission in active jet states suggests a
connection to the ``EUV deficit" of radio loud quasars: radio loud
quasars show a tendency to have a deficit of extreme ultraviolet
(EUV) flux shortward of the peak of the SED around $1100\, \AA$
relative to the EUV flux levels found in radio quiet quasars
\citep{zhe97,tel02,pun14,pun15,pun16}. The magnitude of the deficit
increases with the strength of the jet \citep{pun15}. The analogy is
not direct because the SED of Mrk\,231 is extremely unusual and not
well understood, especially in the UV \citep{vei13,vei16}. The
continuum is likely observed through multiple ``exotic" absorbing
media that differ from the Galactic interstellar medium
\citep{rey12,vei13}. The far UV flux is highly depressed in overall
magnitude, yet the SED is very flat from $1400\,\AA$ to $2000 \,
\AA$, not typical of absorption by a dusty interstellar medium (see
Figure 12). However, we can draw one commonality with the EUV
deficit of radio loud quasars, the flux is lower at wavelengths
shortward of the far UV peak of the SED when there is an active jet
as compared to the flux level when the jet is in a quiescent state.
It might be that the disk in Mrk\,231 is very cool as a consequence
of energy losses due to winds driven by the continuum that is
emitted from the inner disk \citep{lao14}. Regardless of the reason
for the low frequency of the SED turnover, the highest frequency
optically thick thermal emission should represent emission from the
innermost accretion disk. This emission should be detected near or
just blueward of the far UV break in the SED \citep{sun89,szu96}.
Figure 12 seems to indicate that the optically thick thermal
emission from the innermost accretion disk (i.e., emission just
shortward of the peak of the SED) is lower when there is a radio jet
similar to the ``EUV deficit" in radio loud quasars. In particular,
the ``EUV deficit in radio loud quasars" has been shown to be
consistent with the predicted trends of a simple magnetically
arrested accretion scenario in which islands of large scale vertical
magnetic flux thread the innermost accretion flow in radio loud
quasars. This distribution of flux is located in the base of the
relativistic jet and at the same time displace the optically thick
gas responsible for the thermal emission from the innermost
accretion disk, thereby reducing the flux beyond the peak of the SED
relative to that of radio quiet quasars \citep{pun14,pun15}.
Applying this dynamic to Mrk~231, the displaced far UV emitting gas
equates to a diminished far UV luminosity and the increased vertical
flux creates a stronger jet. This creates an inverse correlation
between the far UV luminosity and jet power.

\par The other possibly very interesting finding is point 5, above.
Recall that the existence of large scale jets and BAL winds are
almost mutually exclusive. The propensity for suppressed large scale
emission increases strongly with BALnicity index
\citep{bec00,bec01}. Figure 12 and Table 2, indicate that Mrk\,231
might be a unique source in which we can see the relativistic jet
and high ionization BAL wind anti-correlation occurring in real
time. This wind suppression is naturally explained by models in
which the BAL wind is launched from the innermost regions of the
accretion flow, including the outer boundary of the central
accretion vortex \citep{pun99,pun00}. In these models, the
propensity of a quasar accretion flow to launch either a BAL wind or
a radio jet is determined by the ratio of radiation pressure to
magnetic pressure of the large scale magnetic field in the central
vortex and the inner edge of the accretion disk. Thus, points 4 and
5 are consistent if the jet and BAL wind are both launched from the
inner accretion disk and accretion vortex. In Mrk\,231, the ratio of
magnetic pressure from magnetic islands and thermal radiation
pressure from optically thick gas vacillates. Sometimes the magnetic
islands are plentiful and a modest relativistic jet is launched
(suppressing the BAL wind and far UV) and at other times the
magnetic islands have diffused away and the spectrum of Mrk\,231
hardens in the far UV and a high ionization wind is launched.
\par One might wonder how the increased X-ray flux during the state
of high jet activity in point 6 is consistent with these other two
findings. The most direct interpretation is that the X-ray excess in
the high radio state is a consequence of inverse Compton emission
from the same electrons that produce the synchrotron emission in the
jet. The spectral index would be slightly steeper than that of the
X-ray power law of the coronal X-rays. Considering the superluminal
speed in point 2, above, one would expect significant Doppler
enhancement of a very weak X-ray source and the resultant flux is
likely highly variable. The other possibility is that the reduced
X-ray flux in the radio low states is a result of an X-ray absorbing
wind \citep{fer15}. When the jet is active, the dynamics of this
X-ray wind launching mechanism (X-ray radiation pressure from the
corona) is disrupted by the magnetic islands. The data are not
sufficient to distinguish between the two possibilities. However, we
show in Section 5, that the difference in the NuSTAR spectrum during
the low radio state compared to the high radio states is well
described by the presence of absorption in a photo-ionized wind.
This is the more intriguing of the two possibilities since it
suggests that the existence of a strong jet suppresses both the UV
absorbing  and the X-ray absorbing winds. The latter is a phenomenon
known to occur in Galactic black hole accreting systems, where jet
production is seen to suppress X-ray winds \citep{nei09}. This could
be the first evidence of a direct analogy between the jet-wind
connection in supermassive black holes and stellar mass black holes.
\subsection{Future Observations} Based on these three important
inferences of our study, it seems that obtaining one or two HST COS
G140L observations in a state of low jet activity would verify if
there is a trend for far UV suppression and if the CIV broad
absorption trough returns. Also, based on Figure 14, it would be
interesting to monitor Mrk~231 in a weak jet state in order to see
if the X-ray emission decreases. A combined NuSTAR and XMM-Newton
(or Chandra) observation could verify if the X-ray driven wind that
was claimed in this low state in \citet{fer15} returns. Furthermore,
longer X-ray observations are required to constrain the X-ray driven
wind in both the high and low radio sates. If VLBA observations are
obtained during another radio flare, it is important to stress the
requirement that the u-v coverage not be compromised by splitting
the observation for scheduling purposes. If the observations are
split, care must be taken to ensure that two unique LST ranges are
attained on the two consecutive days. It is far better to delay the
observation a week or two than compromise the u-v coverage. With
this constraint, a multi-epoch campaign should see an ejected
component evolve in space and time. A HST COS G140L observation in
the the rising high flare state has not been obtained and it would
be of interest to see if the spectrum is the same as other times
during the jet formation.

\section{Acknowledgments}
This research has made use of data obtained with NuSTAR, a project
led by Caltech, funded by NASA and managed by NASA/JPL, and has
utilized the NUSTARDAS software package, jointly developed by the
ASDC (Italy) and Caltech (USA). We also made use of data obtained
with XMM-Newton, an ESA science mission with instruments and
contributions directly funded by ESA Member States. This work made
use of Director's Discretionary Time on both NuSTAR and XMM-Newton
for which we thank Fiona Harrison and Norbert Schartel for
approving, as well as the NuSTAR and XMM-Newton SOC for implementing
and coordinating the observations. We would like to thank S.
Veilleux and M. Melendez for sharing their G130M data. The National
Radio Astronomy Observatory is a facility of the National Science
Foundation operated under cooperative agreement by Associated
Universities, Inc. This research has made use of NASA's Astrophysics
Data System Bibliographic Services.
This work made use of the Swinburne University of Technology software correlator, developed as part of the Australian Major National Research Facilities Programme and operated under licence.

\end{document}